%% file: main.tex
\documentclass[
    aps,prb,twocolumn,
	groupedaddress,superscriptaddress,
	amsfonts,amssymb,amsmath,
	citeautoscript,longbibliography,
	letterpaper, nofootinbib
	]{revtex4-2}

\input{main-setup}

\graphicspath{{figures/}}
\setboolean{togglecomments}{true}
\setboolean{toggletodos}{true}
\setboolean{togglechanges}{true}
\usepackage[dvipsnames]{xcolor}
\usepackage[caption=false]{subfig}
\begin{document}
\title{Read--Rezayi fractional Chern insulators in modulated Bernal graphene}

\author{Ruth Mora--Soto$^\bigstar$}
\email{ruthms@mit.edu}
\affiliation{Department of Physics, Massachusetts Institute of Technology, Cambridge, Massachusetts 02139, USA}
\affiliation{Universitat Polit\`ecnica de Catalunya, Barcelona 08034, Spain}

\author{André Grossi Fonseca$^\bigstar$}
\email{agfons@mit.edu}
\affiliation{Department of Physics, Massachusetts Institute of Technology, Cambridge, Massachusetts 02139, USA}
\affiliation{The NSF Institute for Artificial Intelligence and Fundamental Interactions}

\author{Raul Perea-Causin
}
\affiliation{Department of Physics, Stockholm University, AlbaNova University Center, 106 91 Stockholm, Sweden}\affiliation{Nordita, KTH Royal Institute of Technology, Stockholm University and Uppsala University, 106 91 Stockholm, Sweden}

\author{Hui Liu
}
\affiliation{Department of Physics, Stockholm University, AlbaNova University Center, 106 91 Stockholm, Sweden}

\author{Emil J. Bergholtz
}
\affiliation{Department of Physics, Stockholm University, AlbaNova University Center, 106 91 Stockholm, Sweden}

\author{Marin Solja\v ci\'c}
\affiliation{Department of Physics, Massachusetts Institute of Technology, Cambridge, Massachusetts 02139, USA}
\affiliation{The NSF Institute for Artificial Intelligence and Fundamental Interactions}
\affiliation{Research Laboratory of Electronics, Massachusetts Institute of Technology, Cambridge, Massachusetts 02139, USA\\
$^\bigstar$ denotes equal contribution}

\begin{abstract}
Fibonacci anyons provide a universal platform for topological quantum computation, and emerge as low-energy excitations in the $\mathbb{Z}_3$ Read--Rezayi phase in the fractional quantum Hall effect. However, realistic microscopic realizations of this phase in the absence of a magnetic field have remained elusive. We study a model of periodically modulated Bernal bilayer graphene with gate-screened Coulomb
interactions. Using the recently developed target-phase optimization method in conjunction with band-projected exact diagonalization, we identify at filling $\nu=3/5$ a region of parameter space whose ground state is consistent with a Read--Rezayi fractional Chern insulator. The partially filled band from which it arises is a part of a two-band complex which mimics geometric aspects of the lowest and first Landau levels, with the ground state at $\nu=1/2$ consistent with the Moore--Read state. Our results suggest that modulated Bernal graphene can realize delicate non-Abelian fractional quantum Hall states at zero magnetic field,  while demonstrating target-phase optimization as a practical route to discovering such phases in realistic, high-dimensional microscopic models.
\end{abstract}
\maketitle 

\section{Introduction}

Fractional Chern insulators (FCIs) provide a route to fractional quantum Hall physics in the absence of an external magnetic field, replacing Landau levels (LLs) by partially filled Bloch bands \cite{Sorensen2005,tang2011high,neupert2011fractional,sheng2011fractional,regnault2011fractional,Parameswaran2013,BergholtzLiu2013}. This route has become experimentally compelling following the prediction \cite{Abouelkomsan2020,Repellin2020,Ledwith2020,Li2021} and observation \cite{Spanton2018Science,Xie2021MATBGFCI,Park2023FQAH,Lu2024Pentalayer,Redekop2024MoTe2} of Abelian FCIs in moir\'e and graphene-based platforms. The next challenge is to realize non-Abelian FCIs, which are of great interest for quantum information, since their non-Abelian anyons can support fault-tolerant operations by braiding \cite{nayak2008nonabelian,kitaev2009topologicalphasesquantumcomputation}. Among them, the fermionic $\mathbb{Z}_3$ Read--Rezayi (RR) state is particularly relevant, as it is the simplest state hosting anyons whose braiding properties enable universal quantum computation \cite{nayak2008nonabelian, read1999beyond}. 

Stabilizing non-Abelian FCIs in microscopic Hamiltonians remains challenging. The ideal parent Hamiltonians of such states contain $k$-body short-range interactions with $k > 2$, whereas the two-body long-range Coulomb interaction predominates in solid-state platforms.
Additionally, the single-particle conditions under which such phases are favored, if any, are poorly understood relative to their Abelian counterparts.
Nonetheless, recent numerics have provided evidence for the Moore--Read state in moiré continuum models~\cite{Reddy2024Minibands, Xu2025MultipleChern, Ahn2024NonAbelian, Chen2025RobustNonAbelian, Reddy2026AntiTopological, Li2026GeneralizedLandau} . 

On the other hand, evidence for the RR state remains largely restricted to conventional large-field fractional quantum Hall systems \cite{RezayiRead2009,Mong2017Fibonacci,Zhu2015NonAbelian, 6rc3-kjhc}, or in toy lattice models \cite{Bernevig2012Emergent, Liu2013NonAbelian, Zhu2014Identifying, Wang2015Fermionic}.  Recently, numerical evidence was shown for the RR FCI in idealized moiré Hamiltonians  \cite{Liu_2025, wan2026tunablemultibandgeometryfractional}, once again highlighting moiré platforms as attractive systems in which to realize zero-field  fractionalization. However, whether a \emph{realistic} microscopic Hamiltonian can stabilize such a state remains an open question.

Here we address this by showing numerical evidence for the RR state in periodically modulated Bernal bilayer graphene, which can host higher-LL-like bands at zero net field \cite{fujimoto2025vortexability, gao2023graphene}. We employ the recently developed target-phase optimization method~\cite{fonseca2026gradientbasedsearchquantumphases, fonseca2026chiralsc} to search for this phase in the high-dimensional parameter space of the model via a combination of exact diagonalization and gradient-free optimization. 
We find evidence for the fermionic $\mathbb{Z}_3$ RR FCI at $\nu=3/5$ and $\nu=2/5$ in this system with a gate-screened Coulomb interaction projected into a Chern band, which we confirm through large-scale exact diagonalization across different system sizes. 
We also show that the phase survives even after including  Hartree--Fock corrections from filled bands below the active band. Moreover, we find that the quantum geometry of the partially filled band mimics aspects of the first LL, providing further evidence for the RR state. These results highlight modulated Bernal graphene as a promising platform for non-Abelian FCIs and greatly expand the scope of target-phase optimization to delicate quantum phases in realistic Hamiltonians.

\begin{figure}
    \centering
    \includegraphics[width=\linewidth]{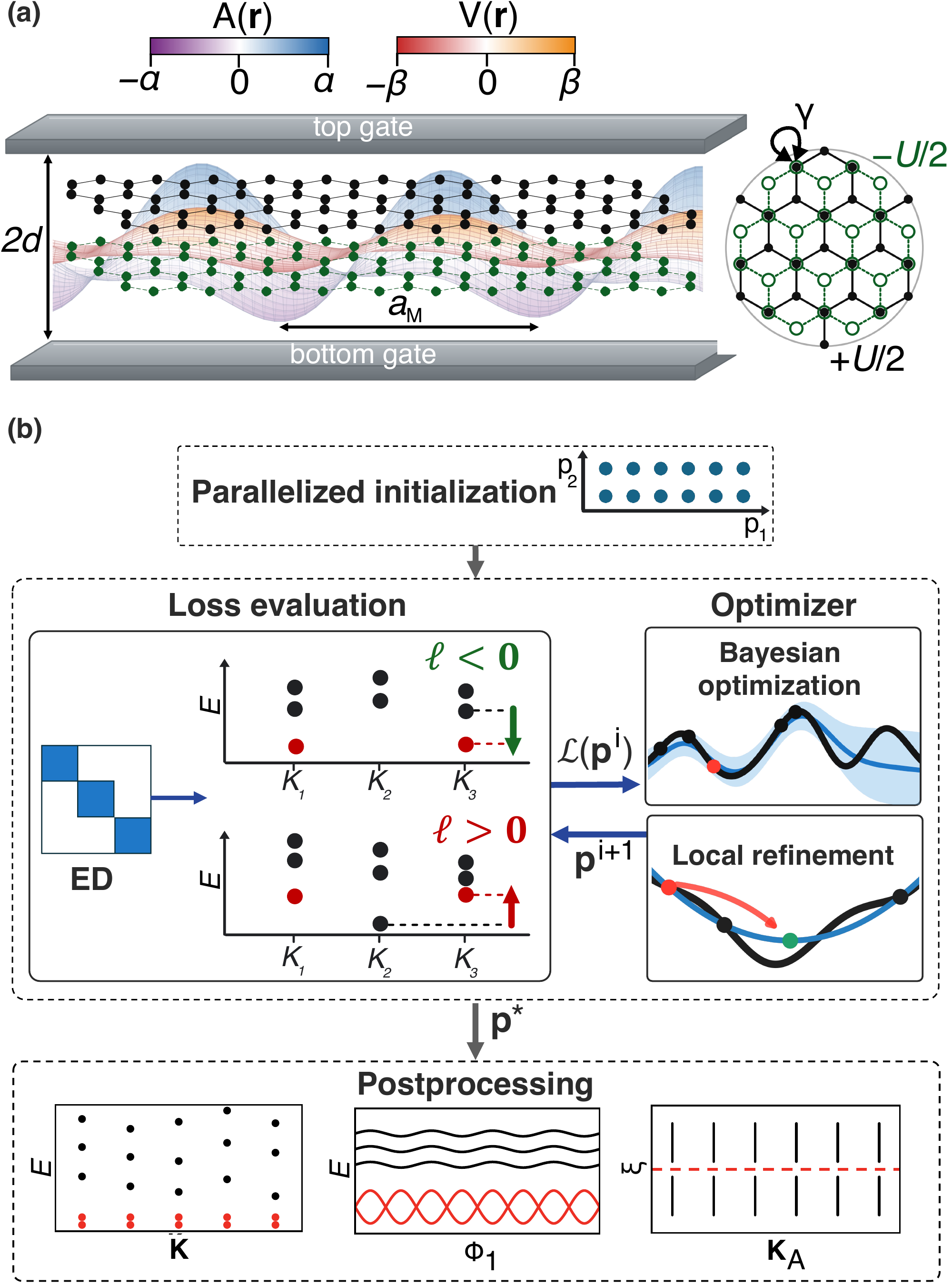}
    \caption{
    (a) Bernal bilayer graphene in the presence of periodic pseudomagnetic and scalar fields (color map) in a dual-gate geometry (b) Schematic workflow for target-phase optimization to search for the Read--Rezayi phase. 
    }
    \label{fig:pipeline_system}
\end{figure}

\section{Modulated Bernal graphene}

We consider spin- and valley-polarized electrons in Bernal bilayer graphene subject to a smooth periodic
strain field and a scalar field, both defined on a triangular superlattice of moiré period $a_M$,
as sketched in \cref{fig:pipeline_system}(a) \cite{fujimoto2025vortexability, gao2023graphene}. Because the modulations vary
slowly on the graphene lattice scale, intervalley scattering is negligible, and therefore we work in the $K$ valley, which is related to the $K'$ valley by time-reversal.

We model the strain-induced pseudomagnetic field of strength $B_0$ by its leading triangular superlattice harmonics,
\begin{equation}
    B(\mathbf r)
    =
    B_0
    \sum_{\ell=0}^{5}
    e^{i\mathbf G_\ell\cdot \mathbf r},
    \qquad
    \mathbf G_\ell
    =
    R_{2\pi\ell/6}\mathbf G_0 ,
\end{equation}
where \(\mathbf G_0=4\pi/\sqrt{3}a_M (1,0)\), and the
\(\mathbf G_\ell\) are the six shortest reciprocal superlattice vectors. The field averages to zero over the superlattice unit cell, so single-particle states form Bloch bands (or pseudo LLs) rather than ordinary LLs. We choose the corresponding dimensionless vector-potential profile
$
    A_x(\mathbf r)+iA_y(\mathbf r)
    =
    A_0 \sum_{\ell=0}^{5}
    e^{i\pi\ell/3}e^{i\mathbf G_\ell\cdot\mathbf r},
$
where $A_0 = B_0/|\mathbf{G}_0|$. We also include a scalar modulation of strength \(V_0\) with the same phase $
    V(\mathbf r)
    =
    V_0
    \sum_{\ell=0}^{5}
    e^{i\mathbf G_\ell\cdot \mathbf r}$.
Then, the single-particle Hamiltonian is
\begin{equation}
\label{eq:hsbg}
H_{\rm sp}
=
E_0
\begin{pmatrix}
h-U/2 & \gamma\,\sigma_+\\
\gamma\,\sigma_- & h+U/2
\end{pmatrix},
\,
h=
\boldsymbol{\sigma}\,\cdot\,
\left[-i\nabla+\alpha\mathbf A(\mathbf r)\right]
+\beta V(\mathbf r)  ,
\end{equation}
where $
E_0=\hbar v_F|\mathbf G_0|,\, \alpha=eB_0/\hbar|\mathbf G_0|^2,\,
\beta=V_0/E_0
$, \(\sigma_\pm=(\sigma_x\pm i\sigma_y)/2\), with
\(v_F\) the graphene Fermi velocity.
The dimensionless parameters $\alpha$, $\beta$, $\gamma$, and $U$
control the strength of the pseudofield, the strength of the scalar potential,
the interlayer tunneling and the interlayer bias, respectively.
We set
\(E_0=300\,\mathrm{meV}\) \cite{mao2020evidence}, which then fixes \(|\mathbf{G}_0|\simeq0.52\,\mathrm{nm}^{-1}\), corresponding to moiré period
\(a_M\simeq14\,\mathrm{nm}\).

We emphasize that all of these ingredients can be experimentally realized. Periodic strain may be generated
using patterned nanorods or substrate-induced buckling
\cite{jiang2017visualizing,mao2020evidence}, while a scalar modulation with
the same superlattice periodicity may be produced using patterned
electrostatic gates or a vertical electric field coupled to the buckling
profile \cite{forsythe2018band,shi2019gate}. The interlayer bias can be controlled by a perpendicular displacement field generated by the metallic gates, which also screen the Coulomb interaction as modeled below.

This set-up is motivated by the fact that, in certain parameter regions, this system can host ``first-vortexable'' band pairs, \ie bands whose quantum geometry can approximate that of the zeroth and
first LLs \cite{fujimoto2025vortexability}. However, vortexability is neither imposed by the optimization nor
assumed in identifying the many-body phase.

We use a realistic repulsive Coulomb interaction screened by symmetric metallic
gates at distances $d$ from the graphene sheet, as shown in \cref{fig:pipeline_system}(a).
This leads to an interaction term of the form\begin{equation}
\label{eq:vgate}
\mathcal{V}(\mathbf q)
=
\frac{e^2}{2\epsilon_0\epsilon_r}
\frac{\tanh(|\mathbf q|d)}{|\mathbf q|},
\end{equation}
where we set the dielectric constant $\epsilon_r = 5$, corresponding to the average dielectric constant of hBN-encapsulated graphene \cite{boronnitride}, and we study the effects of tuning $\epsilon_r$ below. The presence of the gates can interpolate between long-range interactions at large $d$ to short-range interactions when $d \rightarrow 0$, the latter of which usually favors fractionalization \cite{PhysRevLett.51.605}.
We measure $d$ in units of $a_M$ and later convert all of the parameters back to physical units.

We diagonalize \cref{eq:hsbg} in a plane-wave basis and project onto the isolated second band above charge neutrality, and later we discuss the effects of the filled band below it.
This leads to the many-body Hamiltonian
\begin{equation}
\label{eq:projected}
H
=
\sum_{\mathbf k}\epsilon(\mathbf k)
c^\dagger_{\mathbf k}c_{\mathbf k}
+
\frac{1}{2A}
\sum_{\mathbf q}\mathcal{V}(\mathbf q)
:\bar\rho_{\mathbf q}\bar\rho_{-\mathbf q}:,
\end{equation}
where $\epsilon(\mathbf k)$ is the dispersion of the active band and $A$ is the system area. The projected density operator is $\bar{\rho}(\mathbf{q}) = \sum_{\mathbf{k}}\Lambda_{\mathbf{k}}(\mathbf{q}) c^\dagger_{\mathbf{k}+\mathbf{q}} c^{\phantom{}}_{\mathbf{k}}$, with form factor $\Lambda_{\mathbf{k}}(\mathbf{q}) = \langle u_{\mathbf{k}+\mathbf{q}}|u_{\mathbf{k}}\rangle$, where $|u_{\mathbf{k}}\rangle$ is the Bloch wave function of the projected band. The form factors contain the quantum geometry of the band and modify the microscopic interaction after band projection.
We use exact diagonalization (ED) to study this Hamiltonian on a torus, and initially work in the flat-band limit by setting $\epsilon(\mathbf k) = 0$ for convenience; later, we restore the bare dispersion and its Hartree--Fock corrections in order to study a more realistic set-up.
The parameter space for this model $\mathbf{p}=(\alpha,\beta,\gamma,U,d)$ is then five-dimensional, in which we carry out our search method.

\section{Target-phase optimization}
Our main goal in this work is to search the space of ground-states of \cref{eq:projected} for the Read--Rezayi FCI.
However, a naive approach is exponentially hard: a uniform
scan of $D$ parameters (here, $D = 5$) at resolution $r$ in some pre-defined parameter-space region requires $r^{D}$ ED calculations, each of which itself is exponentially expensive on the system size.
Therefore, brute-force grid search becomes quickly unfeasible, even for moderately small $D$.

Instead, we
recast this search problem as optimization, following the recently introduced target-phase optimization perspective~\cite{fonseca2026gradientbasedsearchquantumphases, fonseca2026chiralsc}.
The idea is as follows: rather than exhaustively searching for a quantum phase, one encodes the finite-size fingerprints of the phase into a ``target-phase loss function'', which is a scalar function constructed to be minimized (often, become negative) when the desired phase is realized. 
We can then search directly for parameter regions that minimize the loss function via any optimization algorithm, treating ED as a black-box solver that outputs the loss function.
Generically, this approach is expected to scale more favorably than grid search, and is flexible enough to target a wide variety of phases and be interfaced with a variety of numerical solvers.

\begin{figure*}[t]
    \centering
    \includegraphics[width=\textwidth]{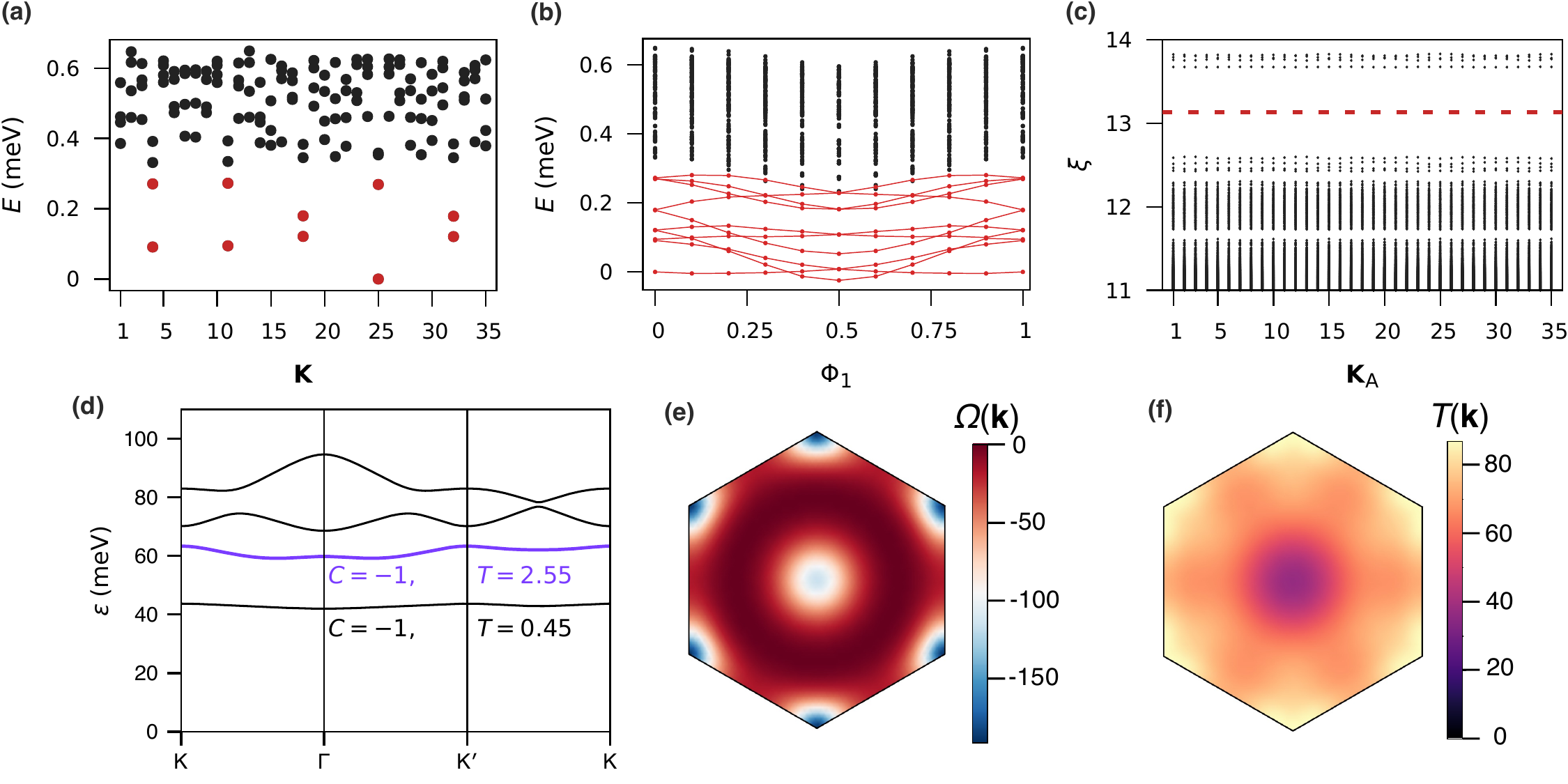}
    \caption{
    Exact-diagonalization signatures of the Read--Rezayi state and quantum geometry at the optimized
    point.
    Results are shown in the flat-band limit for $N_s=35$ sites and $N_e=21$ electrons, at parameter point  $(\alpha, \beta, \gamma, U, d) =(0.636, 0.007, 0.986, 1.033, 2.301)$.
    (a) Low-energy ED spectrum as a function of center-of-mass momentum, with the ground-state manifold highlighted. (b) ED spectrum as a function of inserted flux $\Phi_1$. (c) Particle-cut entanglement spectrum for $N_A = 4$. The number of eigenvalues below red line is 52220, matching exactly the expected RR quasihole count. (d) Band structure along a high symmetry line, with active band highlighted in purple. Chern numbers $C$ and trace violations $T$ are indicated. (e) Berry curvature $\Omega(\mathbf k)$ of the active band. (f) Momentum-resolved trace violation $T(\mathbf k)$ of the active band.} 
    \label{fig:evidence}
\end{figure*}

We now discuss how to construct a target-phase loss function for the RR state.
On a torus, the fermionic $\mathbb{Z}_3$  Read--Rezayi state has a tenfold
ground-state degeneracy, located in many-body momentum sectors $\mathbf{K}_i^*$ with corresponding degeneracies $d_i^*$, which are fixed
by the generalized Pauli principle
\cite{regnault2011fractional,Ardonne_2008,Bernevig_2008,BernevigRegnault2012}.
It is also useful to consider flux insertion through the two
non-contractible cycles of the torus \(\mathbf{\Phi}=(\Phi_1,\Phi_2)\), which is equivalent to imposing twisted boundary
conditions along each direction. 
For a gapped topological phase, adiabatic insertion of one electronic flux
quantum, corresponding to a \(2\pi\) boundary twist, acts as a
unitary transformation within the ground-state manifold, so the many-body gap should remain open throughout the adiabatic cycle \cite{PhysRevB.23.5632, PhysRevLett.84.1535}. 
We use this momentum-resolved ground-state degeneracy and gap resilience to boundary twists to construct a target-phase loss function for the RR state, following \cite{fonseca2026gradientbasedsearchquantumphases}. At a parameter point $\mathbf p$ and inserted flux $\mathbf{\Phi}$, ED yields low-energy levels resolved by momentum $\mathbf{K}$ as a function of the threaded flux, $E_\mathbf{K}(\mathbf{p}, \mathbf{\Phi}$).
We first define the target manifold
$\mathcal{M}_{\mathrm{targ}}$ as the set of $d_i^*$ lowest energy levels in the
predicted momentum sectors $\mathbf{K}_i^*$, and the complement manifold $\mathcal{M}_{\mathrm{comp}}$ as all other energy levels.
Then, the
per-flux loss function is
\begin{equation}
\label{eq:perflux}
\ell(\mathbf{p},\mathbf{\Phi})
= \max_{ E\in\mathcal{M}_{\mathrm{targ}}} E_\mathbf{K}(\mathbf{p},\mathbf \Phi)
- \min_{E\in\mathcal{M}_{\mathrm{comp}}} E_\mathbf{K}(\mathbf{p},\mathbf \Phi).
\end{equation}
By construction, $\ell(\mathbf{p}, \mathbf \Phi)$ is negative in regions of parameter space for which, at a fixed flux, the low energy spectrum exhibits the expected pattern of ground-state degeneracy, with $-\ell$ measuring the many-body gap to higher levels. Otherwise, $\ell(\mathbf{p}, \mathbf \Phi)$ roughly captures a spectral distance in parameter space to the target phase, as depicted in \cref{fig:pipeline_system}(b). Finally, to enforce stability of the many-body gap under
spectral flow, we take the target-phase loss function to be the worst value of per-flux losses over a set $\mathcal{F}$ of fluxes,
\begin{equation}
\label{eq:loss}
\mathcal{L}(\mathbf{p})
= \max_{\mathbf \Phi\in\mathcal{F}} \ell(\mathbf{p},\mathbf \Phi),
\end{equation}
so that $\mathcal{L}<0$ indicates that the ground-state manifold remains spectrally isolated along the flux path.

Our workflow for target-phase search is summarized in \cref{fig:pipeline_system}(b), which we outline here and detail further in the Supplemental Material (SM).
We begin by generating initial seeds in parameter space on a coarse grid, which are optimized in parallel.
At each iteration of the optimization, we carry out an ED calculation to evaluate the loss function, which gets fed through an optimizer.
The optimization is done through a two-stage gradient-free strategy: a global Bayesian optimization stage first explores the parameter space and identifies promising regions with small or negative loss function, and then a local derivative-free refinement further optimizes $\mathcal{L}$ to find the local point with the largest many-body gap.
Our choice for gradient-free optimization stems from the difficulty of computing gradients in our context: finite-differences is numerically expensive, whereas automatic differentiation leads to a large memory overhead on top of the already stringent ED memory requirements. 
Because a negative loss is insufficient to unambiguously identify the ground state, the final step of the workflow is to pass candidates found with $\mathcal{L} < 0$ through a post-processing stage, which consists of stricter diagnostics to distinguish the true RR phase from competing orders and finite-size effects. The diagnostics include inspecting the many-body spectrum at larger system sizes, threading flux and checking for spectral flow of the ground-state manifold, and checking for a gap in the particle-cut entanglement spectrum and the level count below it~\cite{regnault2011fractional, BernevigRegnault2012, Sterdyniak_2011, Sterdyniak_2012, Liu_2025}.

We carry out the searches on a cluster with $N_s = 15$ sites and $N_e = 9$ electrons (see SM for all the clusters used in this work). For this search cluster, the momentum sectors and corresponding degeneracies of the ten ground states should be located at
$
\{(\mathcal K_i^\ast,d_i^\ast)\}
=
\{(1,2),(4,2),(7,2),(10,2),(13,2)\}
$~\cite{Ardonne_2008,Bernevig_2008,BernevigRegnault2012}. 
The sampled fluxes are $\mathcal{F} = \{(0,0),(4\pi/5,0),  (0,4\pi/5)\}$. We choose fluxes $4\pi/5$ since it lies near $\pi$ flux, where finite-size spectral-flow gaps are often smallest, and is commensurate with the filling denominator $\nu=3/5$.

\section{Results}
\cref{fig:evidence} shows ED evidence for a fermionic
$\mathbb{Z}_3$ Read--Rezayi FCI on a $35$-site cluster at $\nu=3/5$, at a parameter point
identified by the optimization. 
In \cref{fig:evidence}(a), we plot the low-energy spectrum as a function of the many-body momentum, which contains ten quasi-degenerate states, separated from the excited spectrum by a many-body gap and located in exactly the  momentum sectors
predicted for the $\mathbb{Z}_3$ RR state by the generalized Pauli counting rules
\cite{read1999beyond,regnault2011fractional,Ardonne_2008,Bernevig_2008,BernevigRegnault2012}.
This tenfold manifold is the finite-size
fingerprint of the expected topological ground-state degeneracy, with a large finite-size splitting that is consistent with previous studies of the RR state~\cite{Liu_2025, wan2026tunablemultibandgeometryfractional}. 
Under flux threading, the ten states
flow among themselves permuting quantum numbers [\cref{fig:evidence}(b)], with weak mixing with excited states along the flux path. This provides further evidence for a fractionalized ground state, with level mixing indicating strong finite-size effects, which we further discuss in the SM. 

A tenfold ground-state manifold in the expected momentum sectors does not by itself establish topological order, since a
competing charge-density wave can often mimic such degeneracy patterns
\cite{LiuLiuBergholtz2024NonAbelian,Wilhelm2021,Abouelkomsan2020}. The sharper diagnostic is the
particle-cut entanglement spectrum, shown in \cref{fig:evidence}(c). We compute it by first partitioning the
$N_e$ particles into $N_A$ and $N_B = N_e - N_A$ particles. For the ten quasi-degenerate ground states, we form the equal-weight density matrix
$\rho=\frac{1}{10}\sum_{i=1}^{10}|\Psi_i\rangle\langle\Psi_i|$ and trace over
the $N_B$ particles to obtain the $N_{A}$-body reduced density matrix 
$\rho_A=\mathrm{Tr}_{N_B}\rho$. We then define the entanglement energies as $\xi_i=-\log \lambda_i$, where $\lambda_i$ are the eigenvalues of $\rho_A$. The entanglement spectrum exhibits a clear entanglement gap as expected, below which the number of levels precisely matches the RR quasihole count \cite{regnault2011fractional, regnault2015entanglementspectroscopyapplicationquantum, Liu_2025} (see SM for a derivation of the analytic counting). This diagnostic thus rules out a symmetry-breaking phase or an Abelian Jain state, which is also competitive at this filling.
In the SM we show that these signatures are reproduced consistently for smaller clusters of $N_s = 15, 20$ and $25$ sites, as well as at filling $\nu=2/5$. However, at $30$ sites the many-body energy gap does not cleanly open; we attribute this to strong finite-size effects, since the RR states remain energetically competitive, and the entanglement gap remains large with the correct count below it across different cluster geometries.
We also show in the SM that the many-body gap closes if the interactions are too short-ranged, which is opposite to the typical intuition for fractional quantum Hall states~\cite{PhysRevLett.51.605}. This behavior also highlights that the interaction profile is crucial to obtain this phase, not only the single-particle quantum geometry.

\begin{figure}[t]
    \centering

    \label{fig:epsilon-gap-refined}
    {
        \includegraphics[width=0.51\textwidth]
        {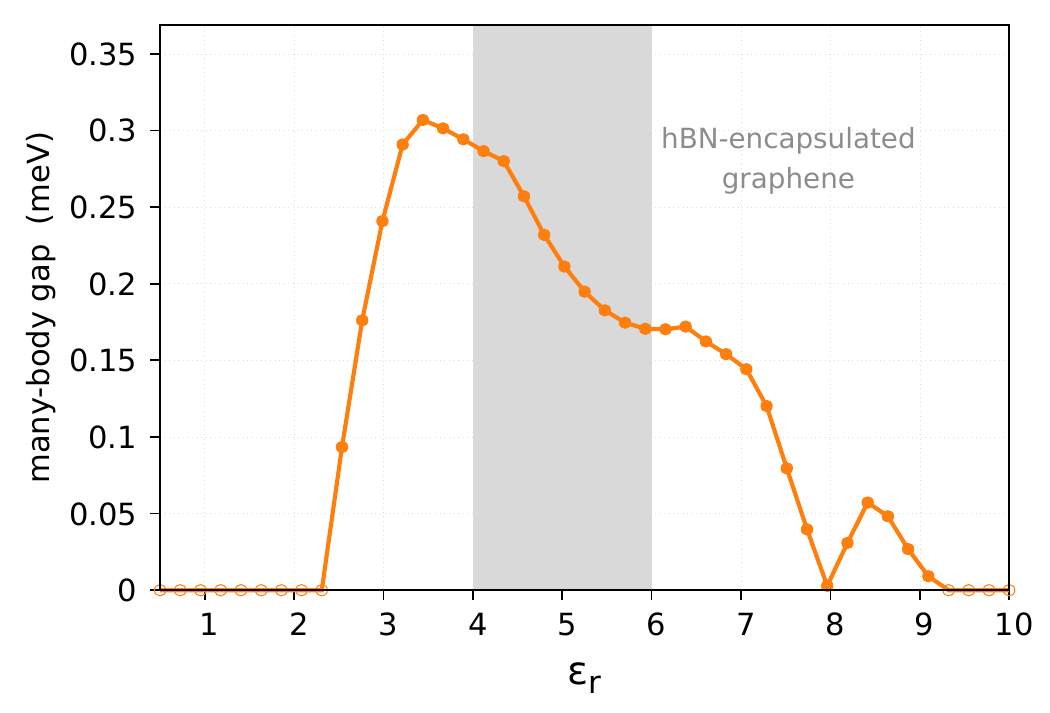}
    }
    \vspace{-5mm}
    \caption{Many-body gap of the Read--Rezayi state in the $N_s = 25$ cluster as a function of dielectric constant $\varepsilon_r$ upon including the bare dispersion and Hartree--Fock self-energy, at parameter point $(\alpha, \beta, \gamma, U, d) =(0.620, 0.006, 0.943, 1.009, 3.110)$. Shaded region denotes the range of dielectric constants of hBN-encapsulated graphene. 
  } 
    \label{fig:epsilon-gap}
\end{figure}

The single-particle data at the optimized point offers further insight into the nature of the many-body state.
\cref{fig:evidence}(d) shows that the active band is accompanied by a band below it, both above charge neutrality and relatively flat.
We now characterize their quantum geometry, which is often crucial in stabilizing fractional quantum Hall states \cite{PhysRevB.90.165139, Simon_2015, Yang_2025, BergholtzLiu2013, fujimoto2025vortexability, PhysRevLett.127.246403, Ledwith2023Vortexability, Liu2026,oriol2026}. In the SM we briefly review concepts in quantum geometry, and define the Berry curvature, Chern number and trace violation, which are most important band quantities in this context. Both bands have Chern number $C=-1$, with the Berry curvature and pointwise trace violation strongly nonuniform (\cref{fig:evidence}(e-f)).  Nevertheless, the trace violations are suggestive of a LL hierarchy: we find $T=2.55$ for the active band and $T=0.45$ for the band below, close to the ideal values $T=2$ and $T=0$ for the first and zeroth LLs, respectively \cite{Ledwith2023Vortexability,fujimoto2025vortexability}. On the other hand, the combined two-band trace violation, $T_{2b}=0.87$, remains somewhat above the ideal vortexable value $T_{2b}=0$~\cite{Ledwith2023Vortexability}. Nonetheless, further strong evidence for the LL-like nature of this two-band complex comes from the observation that, at $\nu=1/2$ of the active band, the ground state is consistent with the Moore--Read state, as we detail in the SM. Taken together, the band geometry indicators and ground state at half filling support a picture in which the two bands approximately mimic the zeroth and first LLs, suggesting a microscopic origin for the Read--Rezayi state observed at $\nu=3/5$.

As noted above, the target-phase optimizations were carried out in the flat-band limit for simplicity. However, this approximation neglects both the bare dispersion of the active band and the static self-energy generated by the completely filled band below it, which can lead to drastically different many-body ground states. We now study such effects by employing a frozen-core Hartree--Fock treatment
\cite{Reddy2024Minibands, Giuliani_Vignale_2005}: the lower band is held fixed as an inert Fermi sea, and its Hartree and Fock self-energies $\Sigma_{\mathrm H}(\mathbf{k}) $ and $\Sigma_{\mathrm F}(\mathbf{k})$ are added to the
active-band dispersion. This leads to a total band dispersion $\tilde{\epsilon}(\mathbf{k}) = \epsilon(\mathbf{k}) +\Sigma_{\mathrm H}(\mathbf{k})+\Sigma_{\mathrm F}(\mathbf{k})$ (see SM for a review).
At the flat-band optimum, including the Hartree--Fock-renormalized one-body dispersion closes the RR gap for the 25-site cluster. However, we run another local refinement on the $15$-site cluster, using the same target-phase
loss function but including the full dispersion during
each loss evaluation. This procedure identifies a nearby parameter point in which the Read--Rezayi manifold is again the ground state, with a finite gap over a broad range of dielectric constants, including the regime of hBN-encapsulated graphene $\epsilon_r\simeq4$--$6$ (\cref{fig:epsilon-gap})~\cite{boronnitride}. Thus, Hartree--Fock corrections shift the optimal stability region rather than eliminate the RR phase. 

We now discuss the experimental scales associated with the RR parameters reported in \cref{fig:epsilon-gap}. For $E_0=300\,\mathrm{meV}$ and $a_M\simeq14\,\mathrm{nm}$, these correspond to a pseudomagnetic field strength $B_0\simeq109\,\mathrm{T}$, scalar potential strength $V_0\simeq1.8\,\mathrm{meV}$, interlayer hopping $\gamma_1\simeq283\,\mathrm{meV}$, and gate distance $d\simeq44\,\mathrm{nm}$. The interlayer bias $U\simeq303\,\mathrm{meV}$ corresponds to a gap $E_g=U\gamma_1/\sqrt{U^2+\gamma_1^2}\simeq207\,\mathrm{meV}$ in the minimal unmodulated tight-binding model for Bernal graphene. The value of $B_0$ needed is accessible in previous realizations~\cite{mao2020evidence}, which reported a pseudomagnetic field amplitude of up to $116\,\mathrm{T}$. The required scalar modulation is modest relative to experimentally demonstrated patterned dielectric architectures~\cite{forsythe2018band, shi2019gate}. Our interlayer hopping is somewhat below the measured value of $\gamma_1 = 380$--$400\,\mathrm{meV}$~\cite{Kuzmenko2009,Zhang2008}; matching this value would require increasing $E_0$ to about $400\,\mathrm{meV}$, corresponding to moiré period $a_M \simeq10\,\mathrm{nm}$, at the expense of stronger pseudomagnetic and scalar modulations. The estimated Bernal graphene gap we obtain from the interlayer bias lies within the experimentally demonstrated gate-tunable range, which extends to approximately $250\,\mathrm{meV}$~\cite{Zhang2009}, while the required gate distance is compatible with hBN thicknesses used in graphene-based quantum Hall devices~\cite{Kim2023,Zibrov2017}. These comparisons motivate experimental exploration of this regime, although the parameter scans in the SM reveal a narrow stability window, indicating that accessing the RR phase would require careful control of the device parameters.
Finally, the values of many-body gaps obtained after Hartree--Fock corrections are similar to those recently reported experimentally for some Abelian states \cite{butler202613fractionalgaplessinteger}, again suggesting a path to experimental observation. 


\section{Discussion}


In this work, we used target-phase optimization to search for and locate a ground state consistent with the fermionic $\mathbb{Z}_3$ Read--Rezayi fractional Chern insulator at filling \(\nu=3/5\), in a realistic model for periodically modulated Bernal bilayer graphene under screened Coulomb interactions.  
We verified a collection of numerical signatures of this state, including ground-state degeneracy, spectral flow and particle-cut entanglement spectrum counts, which distinguish the observed phase from competing charge-ordered and Abelian states. We further find that the state is supported by a two-band complex whose quantum geometry resembles aspects of the lowest and first Landau levels, and that the Read--Rezayi phase survives upon inclusion of the band dispersion and Hartree--Fock self-energy. Our results therefore point to a microscopically motivated platform that supports non-Abelian fractionalization.

An immediate direction for future work is to establish the extent and thermodynamic stability of this phase using larger-scale numerics, such as density-matrix renormalization group calculations and neural quantum states, which would be crucial to resolve finite-size and band mixing effects.
Encouragingly, recent multiband ED studies have indicated that the RR state can be resilient to band mixing ~\cite{wan2026tunablemultibandgeometryfractional}, but a detailed numerical analysis in our platform is needed. Additionally, the appearance of both Read--Rezayi and Moore--Read states in the same band complex suggests a rich phase diagram which needs to be more thoroughly explored, in particular if there are any ways in which it depart from traditional LL phenomenology.

More broadly, our results suggest a systematic route toward engineering Fibonacci anyons in zero-field electronic systems. The combination of periodic strain, electrostatic modulation, interlayer bias, and gate-controlled screening provides several experimentally tunable handles with which to approach the regime we found, while target-phase optimization offers a practical framework for incorporating these experimental constraints directly into the search. Extending this strategy to other systems and target phases points to a shift in quantum-phase search from a largely intuition-driven process into a controlled inverse-design problem. In this sense, our work demonstrates a path to uncover delicate quantum phases systematically within a high-dimensional experimental parameter space.

\section{Acknowledgments}

We thank Patrick J. Ledwith and Raymond Ashoori for stimulating
discussions, and thank Guangyue Ji for independently validating our calculations.
R.M.S. acknowledges the CFIS Mobility Program for its support.
A.G.F. and M.S. acknowledge support from the National Science Foundation under Cooperative Agreement PHY-2019786 (The NSF AI Institute for Artificial Intelligence and Fundamental Interactions), and the U.S.\ Office of Naval Research (ONR) Multidisciplinary University Research Initiative (MURI) under Grant No.\ N00014-20-1-2325 on Robust Photonic Materials with Higher-Order Topological Protection.
This material is based upon work also supported in part by the U. S. Army Research Office through the Institute for Soldier Nanotechnologies at MIT, under Collaborative Agreement Number W911NF-23-2-0121.
The MIT SuperCloud and Lincoln Laboratory Supercomputing Center provided computing resources that contributed to the results reported in this work. 
Any use of generative AI in this manuscript adheres to ethical guidelines for use and acknowledgement of generative AI in academic research. 
Each author has made a substantial contribution to the work, which has been thoroughly vetted for accuracy, and assumes responsibility for the integrity of their contributions~\cite{mann2024ai}.

\bibliography{references}
\end{document}


\raggedbottom

\title{\texorpdfstring{
        SUPPLEMENTAL MATERIAL\\[1ex]
        Read--Rezayi fractional Chern insulators in modulated Bernal graphene
        }
        {Supplemental Material}
       }

\author{Ruth Mora--Soto$^\bigstar$}
\email{ruthms@mit.edu}
\affiliation{Department of Physics, Massachusetts Institute of Technology, Cambridge, Massachusetts 02139, USA}
\affiliation{Universitat Polit\`ecnica de Catalunya, Barcelona 08034, Spain}

\author{André Grossi Fonseca$^\bigstar$}
\email{agfons@mit.edu}
\affiliation{Department of Physics, Massachusetts Institute of Technology, Cambridge, Massachusetts 02139, USA}
\affiliation{The NSF Institute for Artificial Intelligence and Fundamental Interactions}

\author{Raul Perea-Causin
}
\affiliation{Department of Physics, Stockholm University, AlbaNova University Center, 106 91 Stockholm, Sweden}\affiliation{Nordita, KTH Royal Institute of Technology, Stockholm University and Uppsala University, 106 91 Stockholm, Sweden}

\author{Hui Liu
}
\affiliation{Department of Physics, Stockholm University, AlbaNova University Center, 106 91 Stockholm, Sweden}

\author{Emil J. Bergholtz
}
\affiliation{Department of Physics, Stockholm University, AlbaNova University Center, 106 91 Stockholm, Sweden}

\author{Marin Solja\v ci\'c}
\affiliation{Department of Physics, Massachusetts Institute of Technology, Cambridge, Massachusetts 02139, USA}
\affiliation{The NSF Institute for Artificial Intelligence and Fundamental Interactions}
\affiliation{Research Laboratory of Electronics, Massachusetts Institute of Technology, Cambridge, Massachusetts 02139, USA\\
$^\bigstar$ denotes equal contribution}

\maketitle

\setlength{\parindent}{0em}
\setlength{\parskip}{.5em}

\noindent{\small\textbf{\textsf{CONTENTS}}}\\ 
\twocolumngrid
\begingroup
    \let\bfseries\relax 
    \deactivateaddvspace 
    \deactivatetocsubsections 
    \makeatletter\@starttoc{toc}\makeatother 
\endgroup
\onecolumngrid

\count\footins = 1000 
\interfootnotelinepenalty=10000 

\section{Target-phase optimization details}

We minimize the target-phase loss function $\mathcal{L}$ using a two-stage gradient-free strategy. In the first stage, we employ Bayesian optimization to coarsely navigate the parameter space using an exploration-exploitation strategy. The loss landscape is modeled as a Gaussian process, \ie as a random function whose correlations between different points in parameter space are specified by a kernel. After each evaluation of $\mathcal{L}$, this model provides both an estimate of the loss at unseen points and an uncertainty in that estimate. The next point is selected using the expected-improvement criterion, which favors points that are either predicted to improve upon the lowest loss found so far (exploitation) or lie in regions with high uncertainty (exploration). We use a Mat\'ern-$5/2$ kernel, whose finite smoothness is suitable for this problem because $\mathcal{L}$ is not smooth everywhere by construction due to the max and min functions, and because the low-energy spectrum can vary rapidly near phase boundaries. In the second stage, the best point found by Bayesian optimization is locally refined using BOBYQA~\cite{powell2009bobyqa}. This method evaluates $\mathcal{L}$ at several nearby points, fits a quadratic approximation to the local loss landscape, and minimizes this approximation within a region whose size is adjusted as the optimization proceeds. Neither method requires derivatives of $\mathcal{L}$.

A gradient-free formulation is convenient in this
setting because each loss evaluation already requires a memory-intensive
many-body diagonalization. Automatic differentiation through dense eigensolvers is too slow and introduces  additional memory overhead, and differentiating through sparse diagonalization is not available in standard off-the-shelf packages such as PyTorch or JAX.

As explained in the main text, many initial seeds are generated and optimized in parallel, in order to reduce sensitivity to initial conditions and more thoroughly explore the parameter space. For each initial seed, we perform \(N_{\rm BO}=100\) Bayesian-optimization evaluations (unless the loss becomes negative before that), followed by \(N_{\rm BOBYQA}=100\) BOBYQA evaluations. Initial points are chosen from the grid $
\alpha_0 \in \{0.3,\,0.65,\,1.0\}, 
\beta_0  \in \{0.2,\,0.6\},
\gamma_0 \in \{0.3,\,1.2,\,2.1,\,3.0\},
U_0      \in \{0.01,\,0.8\}, 
d_0      \in \{0.01,\,0.1,\,0.3,\,1.0,\,1.5,\,2.0\},$
yielding \(288\) independent optimization runs. 

For the local refinement of the flat-band RR point after the inclusion of the Hartree--Fock self energy, we just use the BOBYQA part of the algorithm 
for a maximum of 100 evaluations maximizing the gap at zero flux, followed by another 100 iterations employing the three fluxes specified in the main text.

\label{sm:optimization}

\section{Finite-size clusters}
\begin{figure}
\centering
\includegraphics[width=\linewidth]{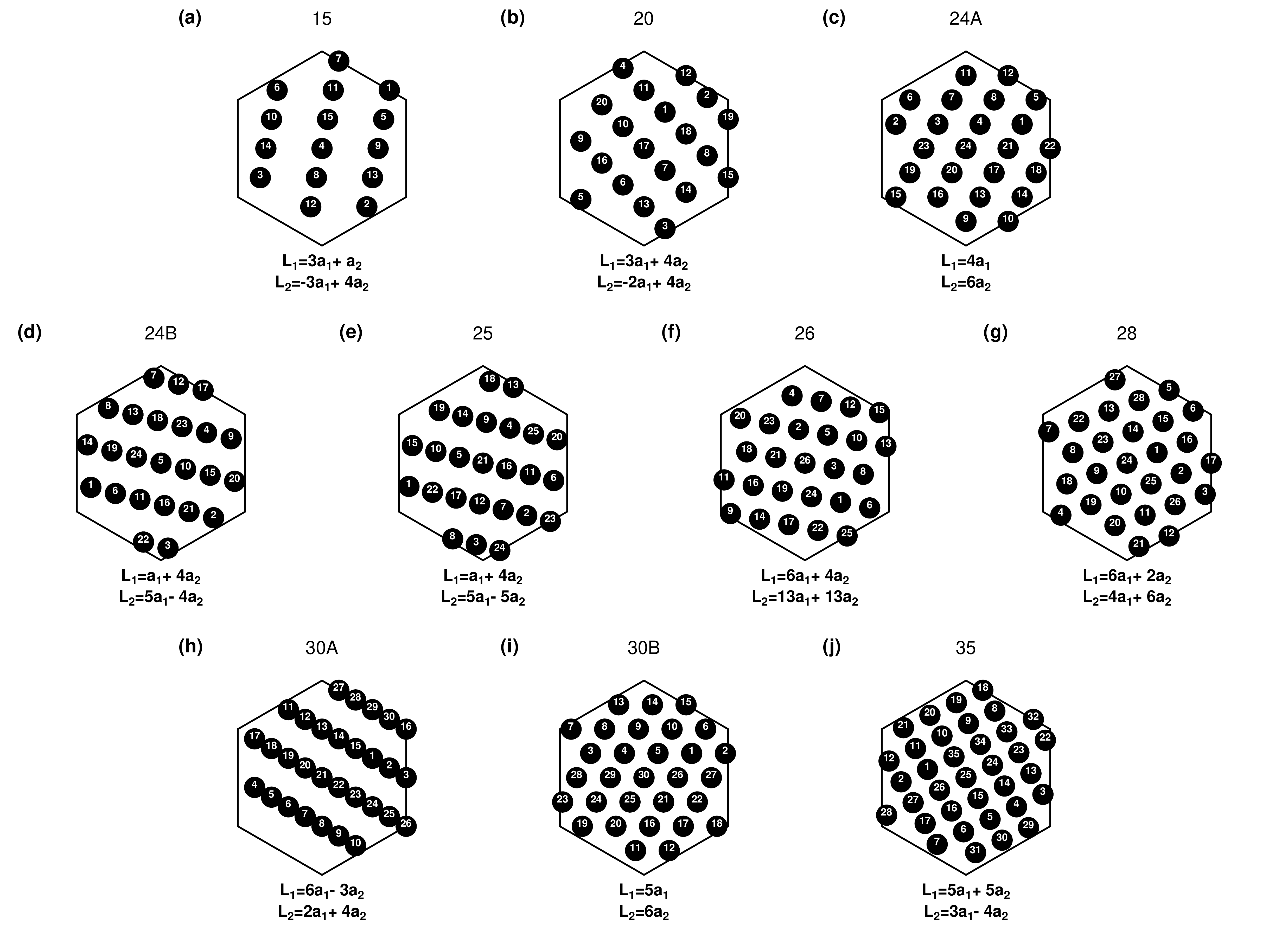}
\caption{Momentum space mesh diagrams of the finite-size clusters used in this work. 
}
\label{fig:momentum_meshes}
\end{figure}

In \cref{fig:momentum_meshes} we plot the allowed momenta of the finite-size clusters used in this work.
$\mathbf{L}_i = n_{xi}\mathbf{a}_1 + n_{yi}\mathbf{a}_2$ are the torus vectors, with
$\mathbf{a}_1 = a_M\big(\tfrac{\sqrt{3}}{2}, -\tfrac{1}{2}\big)$ and $\mathbf{a}_2 = a_M(0, 1)$
dual to the reciprocal lattice vectors $\mathbf{g}_1 = g_0(1, 0)$ and $\mathbf{g}_2 = g_0\big(\tfrac{1}{2}, \tfrac{\sqrt{3}}{2}\big)$.
This geometry leads to $N_s = |n_{x1}n_{y2} - n_{x2}n_{y1}|$ momenta in the mesh, given by $\mathbf{k} = (m_1\mathbf{g}_1 + m_2\mathbf{g}_2)/N$,
with $(m_1, m_2) = k_1(n_{y2}, -n_{x2}) + k_2(-n_{y1}, n_{x1}) \bmod N_s$,
which are folded into the first Brillouin zone and labeled by $k = k_1 + N_1(k_2 - 1)$,
where $k_i = 1, \dots, N_i$, $N_2 = \gcd(N, n_{x2}, n_{y2})$ and $N_1 = N_s/N_2$.

\label{sm:clusters}

\section{Particle-cut entanglement spectra quasihole counting}

For the particle-cut entanglement spectrum (PES), keeping $N_A$ particles in $N_s$ orbitals, the RR quasihole count \cite{Bernevig2012Emergent} is the number of cyclic binary strings with $N_A$ ones satisfying the fermionic $\mathbb Z_3$ RR rule: no more than three particles may occupy any five consecutive orbitals. Equivalently, if $g_j$ is the number of empty orbitals after particle $j$, then
\[
\sum_{j=1}^{N_A}g_j=N_s-N_A,\qquad
g_j+g_{j+1}+g_{j+2}\geq 2
\]
for all $j$, with cyclic indices
\cref{tab:pes-counting} summarizes the resulting PES counts below the gap for the RR state at different system sizes and particle cuts.

\begin{table}[h]
\centering
\begin{tabular}{c c c}
\toprule
$N_s$ & $N_A$ & $\mathcal N_{\rm RR}$ \\
\midrule
15 & 4 & 1305  \\
15 & 5 & 2478  \\
20 & 4 & 4765 \\
20 & 5 & 14404 \\
25 & 4 & 12550  \\
25 & 5 & 51255 \\
30 & 4 & 27285  \\
30 & 5 & 139656 \\
35 & 4 & 52220 \\
\bottomrule
\end{tabular}
\caption{Total quasihole counts for the Read--Rezayi state $\mathcal N_{\rm RR}$ used in the particle-cut entanglement spectra for different cluster sizes $N_s$ and particle cuts $N_A$.}
\label{tab:pes-counting}
\end{table}

\FloatBarrier

The Abelian Jain state has a five-fold ground-state degeneracy in a subset of the momentum sectors of the RR state, and both states occur at filling $\nu = 3/5$. Consequently, finite-size spectra alone can sometimes be difficult to distinguish between these phases. The PES provides an additional diagnostic through its quasihole counting. In all system sizes considered, we observe a clear PES gap at the RR counting, while no other gaps are observed elsewhere across system sizes and particle cuts.

\section{Additional Read--Rezayi evidence}
    \subsection{Additional system sizes at $\nu=3/5$}

\begin{figure}
\centering
\includegraphics[width=\linewidth]{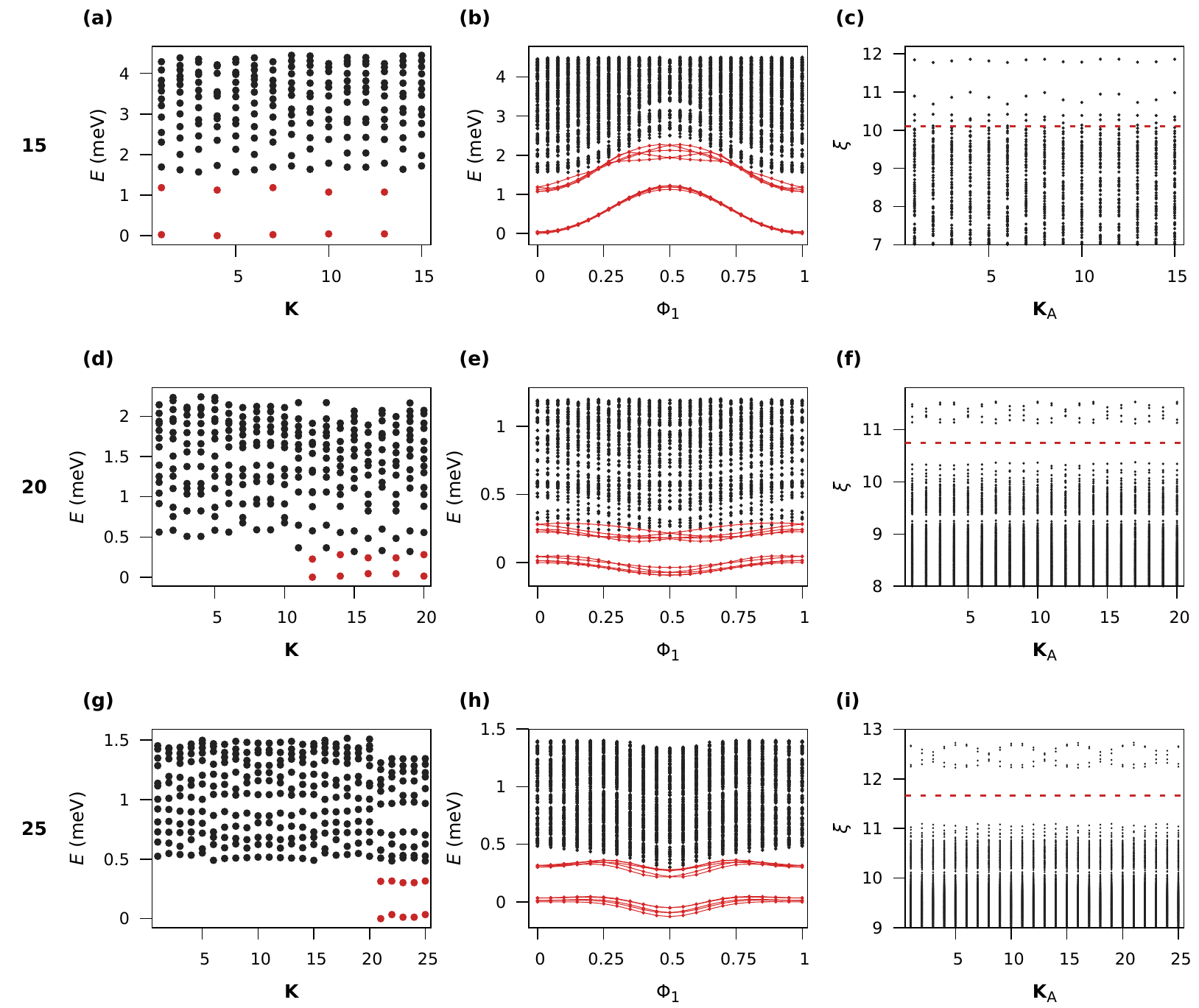}
\caption{Flat-band ED evidence for different clusters at $\nu = 3/5$ and parameters $(\alpha, \beta, \gamma, U, d) =(0.636, 0.007, 0.986, 1.033, 2.301)$ and $\epsilon_r = 5$. 
Data in (a-c) are for $N_s = 15$ cluster, (d-f) are for $N_s = 20$ and (g-i) are for $N_s = 25$. 
(a,d) Low-energy ED spectrum as a function of center-of-mass momentum, with the ground-state manifold highlighted. (b,e) ED spectrum as a function of inserted flux $\Phi_1$. (c,f) Particle-cut entanglement spectrum for $N_A = 4$. The red line marks the expected gap position according to RR quasihole count. 
}
\label{fig:smaller_rr}
\end{figure}

The main text shows evidence for the RR ground state for a $N_s=35$ cluster.~\cref{fig:smaller_rr} shows the corresponding evidence for smaller clusters. For $N_s=15, 20$ and $25$ clusters we see ground-state quasi-degeneracy compatible with the RR state, with once again a large spread. Flux insertion calculations show the expected spectral flow, with the data on the $N_s=20$ cluster weakly mixing with the excited states, although there are no level crossings within the RR momentum sectors. Finally, the PES shows a clean gap for $N_s=20$ and $N_s = 25$, with count below which exactly matching the derived counts from \cref{tab:pes-counting}, while for $N_s=15$ the system is not big enough to show a clear gap. However, for all the computationally feasible systems sizes bigger than $N_s=15$ the PES gap remains clearly open.

\begin{figure}
\centering
\includegraphics[width=0.9\linewidth]{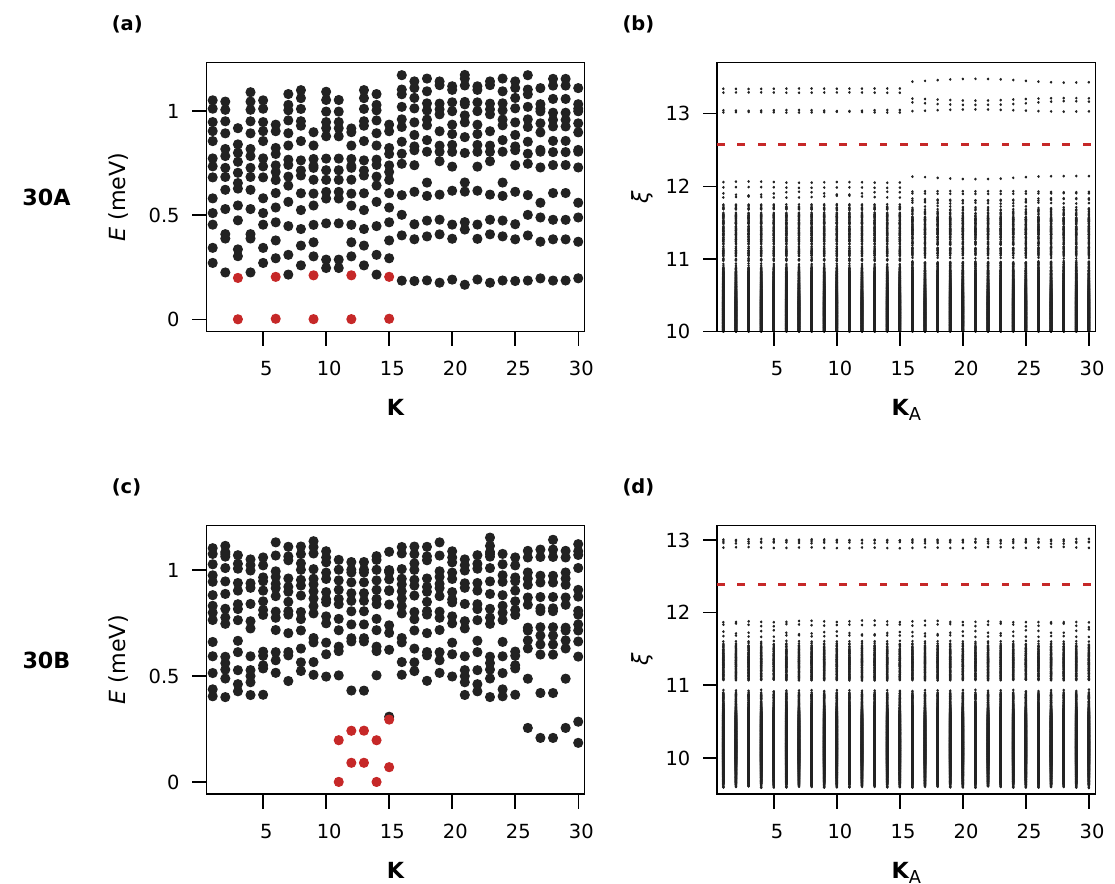}
\caption{
Flat-band ED spectra and PES at $N_A=4$ for 30-site clusters at $\nu = 3/5$ and parameters $(\alpha, \beta, \gamma, U, d) =(0.636, 0.007, 0.986, 1.033, 2.301)$. The number of eigenvalues in the PES below the red line matches exactly the RR expected quasihole counting.
}
\label{fig:30_site_RR}
\end{figure}

However, the story is different for $N_s=30$ clusters, as shown in \cref{fig:30_site_RR}. We find that there is no clean gap above the ten RR states, with the upper five levels slightly mixing with excited states across two different cluster geometries. 
However, we still see a clear, large gap in the PES for both clusters, providing evidence of the RR state as an energetically competitive ground state, but subject to strong finite-size effects. Furthermore, given the sensitivity of the gap to small changes in parameters (see Section S8), it is plausible that there are nearby points in parameter space for which the $N_s = 30$ gap for the RR phase is open, although thoroughly searching for it is beyond our current computational capabilities.

\subsection{Read--Rezayi evidence at $\nu=2/5$}

\begin{figure}
\centering
\includegraphics[width=\linewidth]{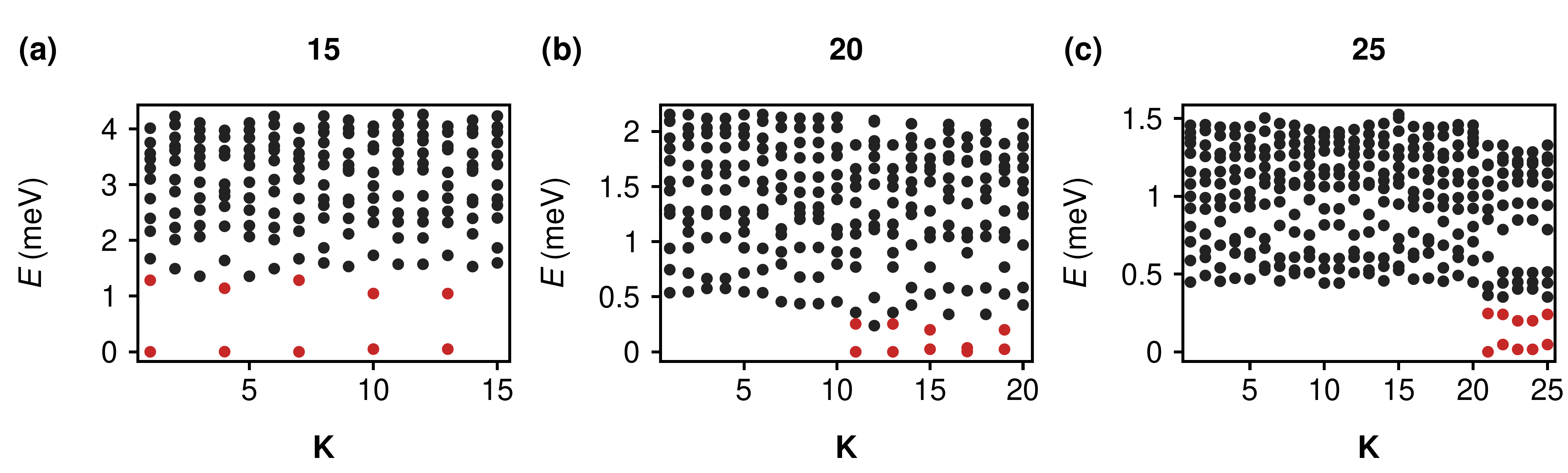}
\caption{
Flat-band ED spectra for different clusters at $\nu = 2/5$ and parameters $(\alpha, \beta, \gamma, U, d) =(0.636, 0.007, 0.986, 1.033, 2.301)$
}
\label{fig:evidence-RR-two-fifths}
\end{figure}

In \cref{fig:evidence-RR-two-fifths} we show spectral evidence for a RR state at the same parameter point found through the flat-band search, but now at the conjugate filling $\nu = 2/5$.
In the partially filled first Landau level the spectrum at these two fillings is exactly the same due to particle-hole symmetry (in the absence of band mixing); here, we find a similar result, albeit particle-hole is at most an approximate symmetry of our system. This provides further evidence for the LL-like nature of the bands and ground states at this parameter point, although there is slight level mixing with excited states on the $N_s = 20$ cluster. 

\section{Quantum geometry and vortexability review}
\label{sm:quantum_geometry}

Here we briefly review concepts in quantum geometry and vortexability, and refer readers to Refs. \cite{PhysRevB.90.165139,Jackson_2015,OzawaMera2021,MeraOzawa2021,PhysRevLett.127.246403,Ledwith2023Vortexability,fujimoto2025vortexability} for more details. For the (isolated) $n$-th band of a translational-invariant system with cell-periodic Bloch state
$|u_n(\mathbf{k})\rangle$, the quantum geometric tensor (QGT) is
\begin{equation}
\mathcal{Q}^{(n)}_{\mu\nu}(\mathbf{k})
=
\langle \partial_{k_\mu}u_n|(1-P_n)|\partial_{k_\nu}u_n\rangle
=
g^{(n)}_{\mu\nu}(\mathbf{k})-
\frac{i}{2}\Omega^{(n)}_{\mu\nu}(\mathbf{k}),
\label{eq:sm_qgt}
\end{equation}
where $P_n=|u_n\rangle\langle u_n|$ is the projector, $g_{\mu\nu}=\operatorname{Re}\mathcal{Q}_{\mu\nu}$ is the Fubini--Study metric, and $\Omega(\mathbf{k})=-2\operatorname{Im}\mathcal{Q}_{xy}$ is the Berry curvature. The integral of the Berry curvature over the Brillouin zone gives the band's Chern number,
$C=(2\pi)^{-1}\int_{\mathrm{BZ}}d^2k\,\Omega(\mathbf{k})$.
Positivity of the QGT implies the following trace inequality:
$\operatorname{Tr}g(\mathbf{k})\geq |\Omega(\mathbf{k})|$. Departures from saturation can be quantified by the momentum-resolved trace violation
\begin{equation}
T(\mathbf{k})=\operatorname{Tr}g(\mathbf{k})-|\Omega(\mathbf{k})|,
\qquad
T=\frac{1}{2\pi}\int_{\mathrm{BZ}}d^2k\,T(\mathbf{k}),
\label{eq:sm_trace_violation}
\end{equation}
where $T$ is refered to as the trace violation. A band is said to be \emph{vortexable} if $T = 0$, which is equivalent to closure of the band under multiplication by the appropriate vortex function, and does not require uniform Berry curvature~\cite{Ledwith2023Vortexability,fujimoto2025vortexability}. For the $n$-th Landau level (nLL), $\operatorname{tr}g_n=(2n+1)|\Omega_n|$, giving $T_n=2n|C|=2n$. Thus, the 0LL and 1LL have $T=0$ and $T=2$, respectively ~\cite{OzawaMera2021}.

A band with $T \simeq 2$ suggests 1LL-like quantum geometry but is, by itself, incomplete information: distinct higher-LL bands can share the same single-band metric data~\cite{fujimoto2025vortexability}. The sharper notion is \emph{first vortexability}. Let $P_1$ project onto the candidate 1LL-like band. It is first vortexable if there exists an orthogonal vortexable partner $P_0$ such that
\begin{equation}
P_0\zeta P_0=\zeta P_0,
\qquad
(P_0+P_1)\zeta P_1=\zeta P_1,
\label{eq:sm_first_vortexable}
\end{equation}
where the chirality of the vortex function $\zeta=x\pm iy$ is chosen according to the sign of the Chern number, and if the two-band subspace is indecomposable into two separately vortexable bands. ~\cref{eq:sm_first_vortexable} expresses a LL ladder structure: multiplying a state in the first band by $\zeta$ may retain it in that band or lower it into its 0LL-like partner, but cannot generate components outside the two-band complex.

This property is naturally tested with a two-band version of the trace violation $T_{\mathrm{2b}}$, with its value being close to zero indicating a nearly first-vortexable two-band complex.
We first define the non-Abelian QGT of the pair,
$\eta^{ab}_{\mu\nu}=\langle\partial_{k_\mu}u_a|(1-P)|\partial_{k_\nu}u_b\rangle$, where $P=P_0+P_1$. The non-Abelian quantum metric and Berry curvature are then $G=\eta_{xx}+\eta_{yy}$ and $F=i(\eta_{xy}-\eta_{yx})$.  The multi-band trace violation is then
$\mathcal{T}=G-\operatorname{sgn}(C)F$. For our two-band complex, this gives $\mathcal{T}=G+F$. Then,
\begin{equation}
T_{\mathrm{2b}}=\frac{1}{2\pi}\int_{\mathrm{BZ}}d^2k\,\operatorname{Tr}\mathcal{T}.
\end{equation}

\section{Moore--Read evidence at $\nu=1/2$}

\begin{figure}
\centering
\includegraphics[width=0.67\linewidth]{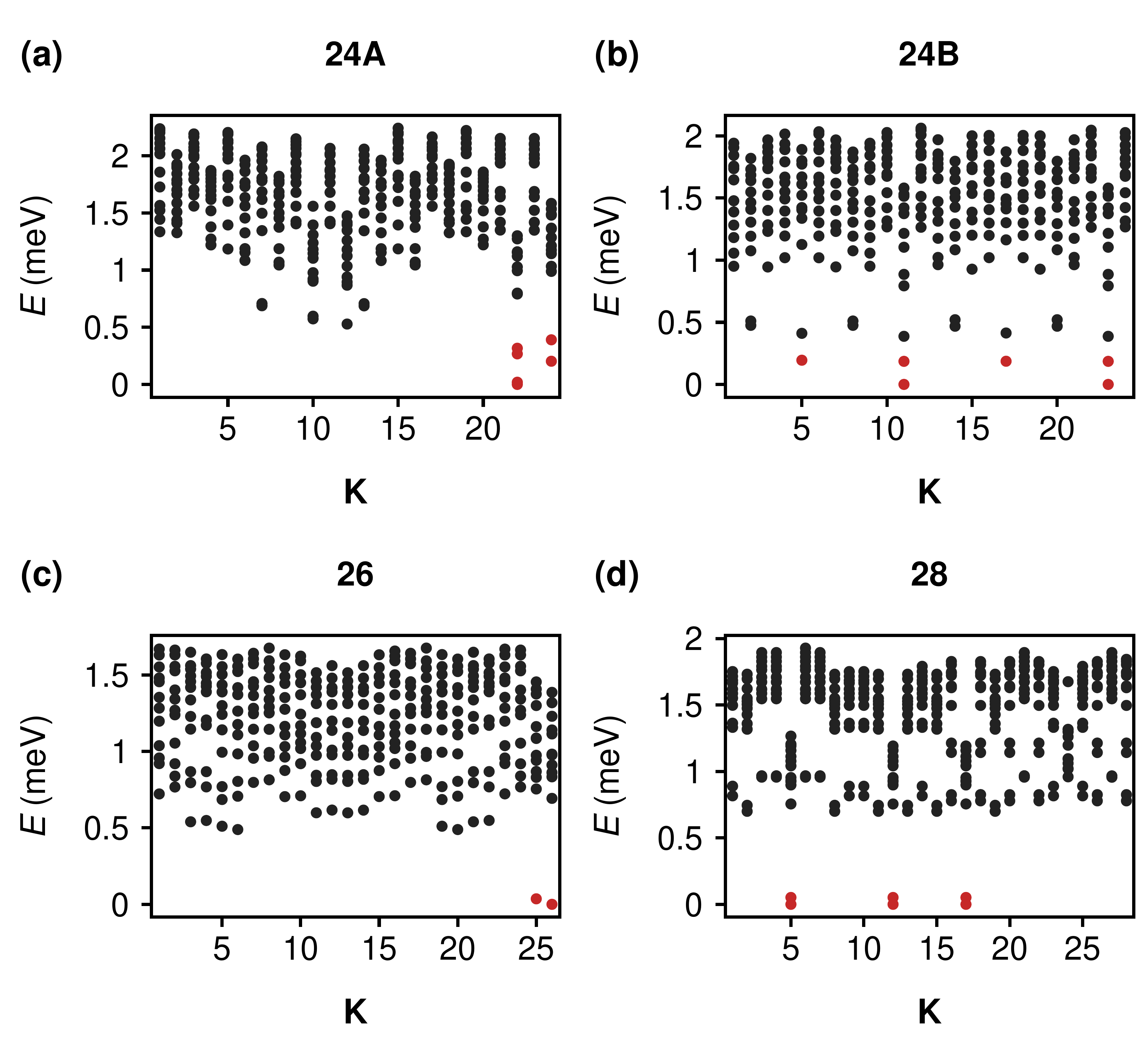}
\caption{
Flat-band ED spectra at $\nu = 1/2$ for $(\alpha, \beta, \gamma, U, d) =(0.636, 0.007, 0.986, 1.033, 2.301)$. The expected ground states for the Moore-Read phase in the corresponding cluster are shown in red.
}
\label{fig:mr}
\end{figure}

In the main text we have argued that, from the viewpoint of quantum geometry, the partially filled active band and the band below it mimic the zeroth and first Landau level, which provides a microscopic origin of the Read--Rezayi state found here.
Here, we show further evidence of LL similarity by showing spectral evidence of the Moore--Read state as the ground state of the active band at filling $\nu = 1/2$, as depicted in \cref{fig:mr}.
We find a ground-state manifold separated by excited states by a clear gap across different system sizes in the predicted momentum sectors, with degeneracy of six/two for even/odd electron number, which is the usual even-odd effect associated to the Moore--Read state~\cite{PhysRevB.61.10267}.

\section{Hartree--Fock self-energy correction}
\label{sm:hf_self_energy}
The exact-diagonalization calculations in the main text are performed in a
single spin- and valley-polarized active band.  In addition to the
active-band dispersion $\epsilon_a(\mathbf k)$ and the interaction projected within that band, we include the static Hartree--Fock potential generated by  the completely filled band immediately below it following \cite{Reddy2024Minibands}. This is a frozen-core approximation, with the filled band kept in its noninteracting Slater determinant and not allowed to fluctuate during the many-body calculation.

Let \(a\) denote the active band and \(f\) the filled band.  Before projection
to the active band, the two-body interaction is
\[
H_{\rm int}
=
\frac{1}{2}
\sum_{ij\ell m}
\sum_{\mathbf{k}'\mathbf{p}'\mathbf{k}\mathbf{p}}
V_{i\mathbf{k}',j\mathbf{p}';\ell\mathbf{k},m\mathbf{p}}\,
c^\dagger_{i\mathbf{k}'}
c^\dagger_{j\mathbf{p}'}
c_{m\mathbf{p}}
c_{\ell\mathbf{k}} .
\]
We normal order this interaction with respect to the Fermi sea
\[
|{\rm FS}\rangle=\prod_{\mathbf p}c^\dagger_{f\mathbf p}|0\rangle .
\]
By Wick's theorem (see Appendix 3 of Ref.~\cite{Giuliani_Vignale_2005}),
the four-fermion operator can be rewritten as a normal-ordered four-fermion
term, one-body terms, and a constant. The one-body terms arise by replacing
creation-annihilation operator pairs by their expectation value in
\(|{\rm FS}\rangle\), e.g.
\(\langle{\rm FS}|c^\dagger_{f\mathbf p}c_{f\mathbf p'}|{\rm FS}\rangle
=\delta_{\mathbf p,\mathbf p'}\).  This gives the frozen-core self energy
\begin{equation}
\Sigma_{ij}(\mathbf k)
=
\sum_{\mathbf p}
\left[
V_{i\mathbf{k},f\mathbf{p};j\mathbf{k},f\mathbf{p}}
-
V_{i\mathbf{k},f\mathbf{p};f\mathbf{p},j\mathbf{k}}
\right].
\end{equation}
The first term is the direct Hartree contribution from the filled-band
density, while the second is the exchange contribution.  Since the calculation
is projected to a single active band, we retain only
\(\Sigma_a(\mathbf k)\equiv\Sigma_{aa}(\mathbf k)\).

Using the generalized form factors
\[
\Lambda^{nm}_{\mathbf k,\mathbf p}(\mathbf q)
=
\langle u_{n\mathbf k}|e^{i\mathbf q\cdot\mathbf r}|u_{m\mathbf p}\rangle ,
\]
the correction is
\[
\Sigma_a(\mathbf k)=\Sigma_{\rm H}(\mathbf k)+\Sigma_{\rm F}(\mathbf k),
\]
with
\[
\Sigma_{\rm H}(\mathbf k)
=
\frac{1}{A}
\sum_{\mathbf G}
\mathcal V(\mathbf G)\,
\Lambda^{aa}_{\mathbf k,\mathbf k}(\mathbf G)\,
\rho_f(-\mathbf G),
\qquad
\rho_f(\mathbf G)
=
\sum_{\mathbf p}
\Lambda^{ff}_{\mathbf p,\mathbf p}(\mathbf G),
\]
and
\[
\Sigma_{\rm F}(\mathbf k)
=
-\frac{1}{A}
\sum_{\mathbf p,\mathbf G}
\mathcal V(\mathbf k-\mathbf p+\mathbf G)\,
\Lambda^{af}_{\mathbf k,\mathbf p}(\mathbf k-\mathbf p+\mathbf G)\,
\Lambda^{fa}_{\mathbf p,\mathbf k}(\mathbf p-\mathbf k-\mathbf G).
\]
Here \(A\) is the total sample area, \(\mathbf G\) runs over reciprocal vectors, and \(\mathcal V(\mathbf q)\) is the screened Coulomb
interaction defined in the main text.

The frozen-core correction enters through the renormalized dispersion
\[
\widetilde{\epsilon}(\mathbf k)
=
\epsilon_a(\mathbf k)+\Sigma_{\rm H}(\mathbf k)+\Sigma_{\rm F}(\mathbf k).
\]
Constant terms produced by normal ordering are omitted because they contribute by a trivial shift to the Hamiltonian at fixed filling.

\section{Read--Rezayi gap stability}
\begin{figure}
\centering
\includegraphics[width=\linewidth]{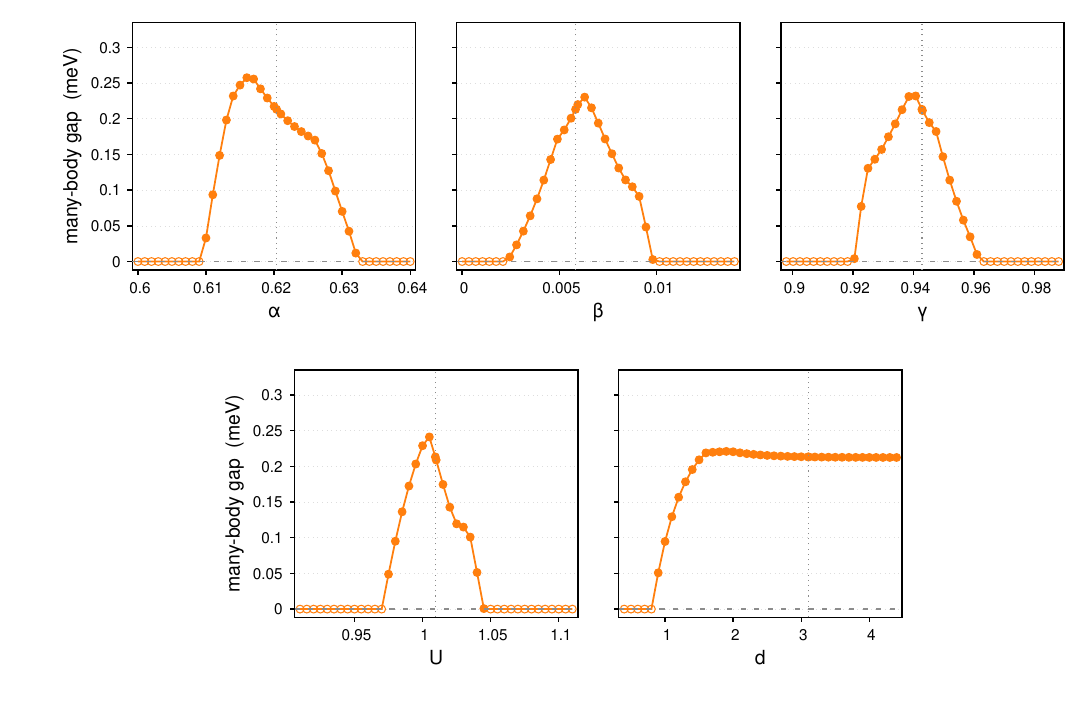}n
\caption{
Many-body gap of the RR state on the $N=25$ cluster as a function of different parameters. Vertical dashed lines indicate the optimized point found after including the full renormalized band dispersion.
}
\label{fig:sweeps-rr}
\end{figure}

An important question for any quantum phase is its stability under parameter perturbations.
In \cref{fig:sweeps-rr} we show the many-body gap of the RR phase including the full dispersion in the 25-site cluster, as we nudge each parameter while fixing all of the others.
We find that the RR state is fairly sensitive to changes in parameters, indicating that the region of stability in parameter space is small.
One noteworthy aspect is that the RR phase is only stabilized when $d$ is sufficiently large, \ie when interactions are long-ranged enough. This goes against the usual expectations for short-range interactions stabilizing fractional quantum Hall states \cite{PhysRevLett.51.605}, and emphasizes that the single-particle quantum geometry is insufficient to determine the many-body ground state, with the interaction profile playing a crucial role here.

\bibliographystyle{apsrev4-2}
\bibliography{references}

%% file: main-setup.tex
\usepackage[utf8]{inputenc}
\usepackage[english]{babel}

\usepackage{microtype} 
\usepackage{xspace} 

\usepackage{txfonts}  
\usepackage{txfontsb} 

\usepackage{bm} 

\usepackage{xcolor}
\usepackage[]{graphicx} 
\graphicspath{{figs/}}

\usepackage[]{booktabs}
\usepackage{array}
\usepackage{layouts}
\usepackage{multirow}

\usepackage{enumerate}
\usepackage[inline]{enumitem}

\usepackage{xr}
\makeatletter
\newcommand*{\addFileDependency}[1]{
  \typeout{(#1)}
  \@addtofilelist{#1}
  \IfFileExists{#1}{}{\typeout{No file #1.}}
}
\makeatother

\usepackage{hyperref}
\hypersetup{colorlinks,
	linkcolor={blue!75!black!80!yellow},
	citecolor={blue!75!black!80!yellow},
	urlcolor={blue!75!black!80!yellow}
}

\usepackage[capitalize,nameinlink]{cleveref}

\crefname{subequations}{Eqs.}{Eqs.} 
\Crefname{subequations}{Eqs.}{Eqs.}
\crefformat{subequations}{#2Eqs.~(#1)#3}
\Crefformat{subequations}{#2Eqs.~(#1)#3}
\crefname{page}{p.}{p.} 
\crefname{table}{Table}{Tables}
\crefname{figure}{Figure}{Figures}
\crefname{section}{Section}{Sections}

\usepackage{placeins}

\usepackage{siunitx}
\DeclareSIUnit[number-unit-product = ]\percent{\char`\%} 

\usepackage[centering,hmargin=18mm,tmargin=29.4mm,bmargin=24mm]{geometry}

\usepackage{soul}

\usepackage{textcomp} 
\usepackage{xifthen}
\usepackage{etoolbox}
\newboolean{togglecomments}
\newboolean{toggletodos}
\newboolean{togglechanges}

\setboolean{togglecomments}{true}
\setboolean{toggletodos}{true}
\setboolean{togglechanges}{false} 

\newcommand{\textblacksquare}{$\blacksquare$}
\newcommand{\todo}[1]{\ifbool{toggletodos}%
	{\textcolor{green!60!black}{\small\textsf{{}\textsuperscript{\textsc{\textsf{todo}}}}[\ignorespaces#1]}} 
	{}}     
\newcommand{\comment}[2]{\ifbool{togglecomments}%
		{\textcolor{blue!70!black}{\small\sf\textsuperscript{\textsc{\textsf{\ignorespaces#1}}}[\ignorespaces#2]}} 
		{}}     

\newcommand{\commentAGF}[1]{\ifbool{togglecomments}%
		{\textcolor{red!70!black}{\small\sf\textsuperscript{\textsc{\textsf{\ignorespaces AGF}}}[\ignorespaces#1]}} 
		{}}     

\newcommand{\commentRMS}[1]{\ifbool{togglecomments}%
		{\textcolor{violet!70!black}{\small\sf\textsuperscript{\textsc{\textsf{\ignorespaces RMS}}}[\ignorespaces#1]}} 
		{}}     

\newcommand{\reply}[2]{\ifbool{togglecomments}%
		{\textcolor{red!70!black}{\small\sf\textsuperscript{\textsc{\textsf{\ignorespaces#1}}}[\ignorespaces#2]}} 
		{}} 
        
\newcommand{\swap}[2]{\ifbool{togglechanges}
	{\ignorespaces#2}  
	{\textcolor{red!70!black}{[\ignorespaces#1]}\textrightarrow{}\textcolor{green!50!black}{[\ignorespaces#2]}}}
\newcommand{\remove}[1]{\ifbool{togglechanges}
	{}    
	{\textcolor{red!70!black}{\ignorespaces#1}}}
\newcommand{\inset}[1]{\ifbool{togglechanges}
	{\ignorespaces#1}  
	{\textcolor{green!50!black}{\ignorespaces#1}}}

\newcommand{\citeremind}[1]{%
	[\textcolor{blue!75!black!80!yellow}{\textblacksquare%
		\ifthenelse{\isempty{#1}}{}{\textsuperscript{\tiny\textsf{\ignorespaces#1}}}%
	}]\xspace}

\input{commands.tex}

\makeatletter
\newcommand{\raisemath}[1]{\mathpalette{\raisem@th{#1}}}
\newcommand{\raisem@th}[3]{\raisebox{#1}{$#2#3$}}
\makeatother

\renewcommand{\paragraph}[1]{\vskip 1ex\noindent\textbf{#1.}~}

\usepackage{braket}
\usepackage[eulergreek]{sansmath}
\makeatletter
\renewcommand\@make@capt@title[2]{%
    \@ifx@empty\float@link{\@firstofone}{\expandafter\href\expandafter{\float@link}}%
    \sisetup{math-sf=\textsf}%
    \sansmath\sffamily\textbf{#1\@caption@fignum@sep}#2 
}%

\makeatother


%% file: commands.tex
\newcommand{\ie}{i.e.,\@\xspace} 

\newcommand{\appropto}{\mathrel{\vcenter{
			\offinterlineskip\halign{\hfil$##$\cr
				\propto\cr\noalign{\kern.2pt}\sim\cr\noalign{\kern-2.5pt}}}}}

\DeclareFontFamily{U}{mathx}{\hyphenchar\font45}
\DeclareFontShape{U}{mathx}{m}{n}{<5> <6> <7> <8> <9> <10>
                                  <10.95> <12> <14.4> <17.28> <20.74> <24.88>
                                  mathx10}{}
\DeclareSymbolFont{mathx}{U}{mathx}{m}{n}
\DeclareFontSubstitution{U}{mathx}{m}{n}

%% file: references.bib
@article{Sorensen2005,
  title = {Fractional Quantum Hall States of Atoms in Optical Lattices},
  author = {S\o{}rensen, Anders S. and Demler, Eugene and Lukin, Mikhail D.},
  journal = {Phys. Rev. Lett.},
  volume = {94},
  issue = {8},
  pages = {086803},
  numpages = {4},
  year = {2005},
  month = {Mar},
  publisher = {American Physical Society},
  doi = {10.1103/PhysRevLett.94.086803},
  url = {https://link.aps.org/doi/10.1103/PhysRevLett.94.086803}
}

@article{Liu2026,
  title = {Topological order without band topology in moir\'e graphene},
  author = {Liu, Hui and Perea-Causin, Raul and Liu, Zhao and Bergholtz, Emil J.},
  journal = {Phys. Rev. Res.},
  volume = {8},
  issue = {3},
  pages = {L032039},
  numpages = {8},
  year = {2026},
  month = {Sep},
  publisher = {American Physical Society},
  doi = {10.1103/jx5w-35z9},
  url = {https://link.aps.org/doi/10.1103/jx5w-35z9}
}

@article{PhysRevB.61.10267,
  title = {Paired states of fermions in two dimensions with breaking of parity and time-reversal symmetries and the fractional quantum Hall effect},
  author = {Read, N. and Green, Dmitry},
  journal = {Phys. Rev. B},
  volume = {61},
  issue = {15},
  pages = {10267--10297},
  numpages = {0},
  year = {2000},
  month = {Apr},
  publisher = {American Physical Society},
  doi = {10.1103/PhysRevB.61.10267},
  url = {https://link.aps.org/doi/10.1103/PhysRevB.61.10267}
}

@article{Li2021,
  title = {Spontaneous fractional Chern insulators in transition metal dichalcogenide moir\'e superlattices},
  author = {Li, Heqiu and Kumar, Umesh and Sun, Kai and Lin, Shi-Zeng},
  journal = {Phys. Rev. Res.},
  volume = {3},
  issue = {3},
  pages = {L032070},
  numpages = {6},
  year = {2021},
  month = {Sep},
  publisher = {American Physical Society},
  doi = {10.1103/PhysRevResearch.3.L032070},
  url = {https://link.aps.org/doi/10.1103/PhysRevResearch.3.L032070}
}

@article{Repellin2020,
  title = {Chern bands of twisted bilayer graphene: Fractional Chern insulators and spin phase transition},
  author = {Repellin, C\'ecile and Senthil, T.},
  journal = {Phys. Rev. Res.},
  volume = {2},
  issue = {2},
  pages = {023238},
  numpages = {8},
  year = {2020},
  month = {May},
  publisher = {American Physical Society},
  doi = {10.1103/PhysRevResearch.2.023238},
  url = {https://link.aps.org/doi/10.1103/PhysRevResearch.2.023238}
}

@article{Ledwith2020,
  title = {Fractional Chern insulator states in twisted bilayer graphene: An analytical approach},
  author = {Ledwith, Patrick J. and Tarnopolsky, Grigory and Khalaf, Eslam and Vishwanath, Ashvin},
  journal = {Phys. Rev. Res.},
  volume = {2},
  issue = {2},
  pages = {023237},
  numpages = {12},
  year = {2020},
  month = {May},
  publisher = {American Physical Society},
  doi = {10.1103/PhysRevResearch.2.023237},
  url = {https://link.aps.org/doi/10.1103/PhysRevResearch.2.023237}
}

@article{Wilhelm2021,
  title = {Interplay of fractional Chern insulator and charge density wave phases in twisted bilayer graphene},
  author = {Wilhelm, Patrick and Lang, Thomas C. and L\"auchli, Andreas M.},
  journal = {Phys. Rev. B},
  volume = {103},
  issue = {12},
  pages = {125406},
  numpages = {16},
  year = {2021},
  month = {Mar},
  publisher = {American Physical Society},
  doi = {10.1103/PhysRevB.103.125406},
  url = {https://link.aps.org/doi/10.1103/PhysRevB.103.125406}
}

@article{gao2023graphene,
  title = {Untwisting Moir\'e Physics: Almost Ideal Bands and Fractional {Chern} Insulators in Periodically Strained Monolayer Graphene},
  author = {Gao, Qiang and Dong, Junkai and Ledwith, Patrick and Parker, Daniel and Khalaf, Eslam},
  journal = {Phys. Rev. Lett.},
  volume = {131},
  issue = {9},
  pages = {096401},
  numpages = {9},
  year = {2023},
  month = {Aug},
  publisher = {American Physical Society},
  doi = {10.1103/PhysRevLett.131.096401},
  url = {https://link.aps.org/doi/10.1103/PhysRevLett.131.096401}
}

@article{fujimoto2025vortexability,
  author       = {Fujimoto, Manato and Parker, Daniel E. and Dong, Junkai and Khalaf, Eslam and Vishwanath, Ashvin and Ledwith, Patrick},
  title        = {Higher Vortexability: Zero-Field Realization of Higher Landau Levels},
  journal = {Physical Review Letters},
  year         = {2025},
  volume       = {134},
  number       = {10},
  eid          = {106502},
  doi          = {10.1103/PhysRevLett.134.106502},
  url          = {https://doi.org/10.1103/PhysRevLett.134.106502},
}

@article{Ledwith2023Vortexability,
  author       = {Ledwith, Patrick J. and Vishwanath, Ashvin and Parker, Daniel E.},
  title        = {Vortexability: A unifying criterion for ideal fractional {Chern} insulators},
  journal      = {Physical Review B},
  year         = {2023},
  volume       = {108},
  number       = {20},
  eid          = {205144},
  doi          = {10.1103/PhysRevB.108.205144},
  url          = {https://doi.org/10.1103/PhysRevB.108.205144},
}

@article{Reddy2024Minibands,
  author       = {Reddy, Aidan P. and Paul, Nisarga and Abouelkomsan, Ahmed and Fu, Liang},
  title        = {Non-Abelian fractionalization in topological minibands},
  journal      = {Physical Review Letters},
  shortjournal = {Phys. Rev. Lett.},
  year         = {2024},
  volume       = {133},
  eid          = {166503},
  doi          = {10.1103/PhysRevLett.133.166503},
  eprinttype   = {arxiv},
  eprintclass  = {cond-mat.mes-hall},
  eprint       = {2403.00059},
  url          = {https://doi.org/10.1103/PhysRevLett.133.166503},
}

@misc{fonseca2026gradientbasedsearchquantumphases,
      title={Gradient-based search of quantum phases: discovering unconventional fractional Chern insulators}, 
      author={André Grossi Fonseca and Eric Wang and Sachin Vaidya and Patrick J. Ledwith and Ashvin Vishwanath and Marin Soljačić},
      year={2026},
      eprint={2509.10438},
      archivePrefix={arXiv},
      primaryClass={cond-mat.str-el},
      url={https://arxiv.org/abs/2509.10438}, 
}

@techreport{Powell2009BOBYQA,
  author      = {Powell, M. J. D.},
  title       = {The {BOBYQA} algorithm for bound constrained optimization without derivatives},
  institution = {Department of Applied Mathematics and Theoretical Physics, University of Cambridge},
  number      = {DAMTP 2009/NA06},
  year        = {2009},
  url         = {https://optimization-online.org/2010/05/2616/},
}

@misc{kitaev2009topologicalphasesquantumcomputation,
      title={Topological phases and quantum computation}, 
      author={Alexei Kitaev and Chris Laumann},
      year={2009},
      eprint={0904.2771},
      archivePrefix={arXiv},
      primaryClass={cond-mat.mes-hall},
      url={https://arxiv.org/abs/0904.2771}, 
}

@article{read1999beyond,
  author       = {Read, N. and Rezayi, E. H.},
  title        = {Beyond paired quantum {Hall} states: Parafermions and incompressible states in the first excited Landau level},
  journal = {Physical Review B},
  year         = {1999},
  volume       = {59},
  number       = {12},
  pages        = {8084--8094},
  doi          = {10.1103/PhysRevB.59.8084},
  url          = {https://doi.org/10.1103/PhysRevB.59.8084},
  eprint       = {cond-mat/9809384},
  eprinttype   = {arxiv},
  
}

@article{nayak2008nonabelian,
  author    = {Nayak, Chetan and Simon, Steven H. and Stern, Ady and Freedman, Michael and {Das Sarma}, Sankar},
  title     = {Non-Abelian anyons and topological quantum computation},
  journal   = {Reviews of Modern Physics},
  issn      = {1539-0756},
  volume    = {80},
  number    = {3},
  pages     = {1083--1159},
  year      = {2008},
  month     = {9},
  publisher = {American Physical Society (APS)},
  doi       = {10.1103/RevModPhys.80.1083},
  url       = {http://dx.doi.org/10.1103/RevModPhys.80.1083},
}

@article{Sterdyniak_2011,
   title={Extracting Excitations from Model State Entanglement},
   volume={106},
   ISSN={1079-7114},
   url={http://dx.doi.org/10.1103/PhysRevLett.106.100405},
   DOI={10.1103/physrevlett.106.100405}, pages={100405},
   number={10},
   journal={Physical Review Letters},
   publisher={American Physical Society (APS)},
   author={Sterdyniak, A. and Regnault, N. and Bernevig, B. A.},
   year={2011},
   month={3} }

@article{Sterdyniak_2012,
   title={Particle entanglement spectra for quantum Hall states on lattices},
   volume={86},
   ISSN={1550-235X},
   url={http://dx.doi.org/10.1103/PhysRevB.86.165314},
   DOI={10.1103/physrevb.86.165314}, pages={165314},
   number={16},
   journal={Physical Review B},
   publisher={American Physical Society (APS)},
   author={Sterdyniak, Antoine and Regnault, Nicolas and Möller, Gunnar},
   year={2012},
   month={10} }

@article{Liu_2025,
  author  = {Liu, Hui and Perea-Causin, Raul and Bergholtz, Emil J.},
  title   = {Parafermions in Moir{\'e} minibands},
  journal = {Nature Communications},
  volume  = {16},
  number  = {1},
  year    = {2025},
  doi={10.1038/s41467-025-57035-x}, pages={1770},
  url     = {https://doi.org/10.1038/s41467-025-57035-x},
}

@article{Spanton2018Science,
  author  = {Spanton, Eric M. and Zibrov, Alexander A. and Zhou, Haoxin and Taniguchi, Takashi and Watanabe, Kenji and Zaletel, Michael P. and Young, Andrea F.},
  title   = {Observation of fractional Chern insulators in a van der {Waals} heterostructure},
  journal = {Science},
  volume  = {360},
  number  = {6386},
  pages   = {62--66},
  year    = {2018},
  doi     = {10.1126/science.aat8459},
  url     = {https://doi.org/10.1126/science.aat8459},
}

@article{tang2011high,
  author  = {Tang, Evelyn and Mei, Jia-Wei and Wen, Xiao-Gang},
  title   = {High-temperature fractional quantum {Hall} states},
  journal = {Phys. Rev. Lett.},
  volume  = {106},
  pages   = {236802},
  year    = {2011},
  doi     = {10.1103/PhysRevLett.106.236802},
}

@article{neupert2011fractional,
  author  = {Neupert, Titus and Santos, Luiz and Chamon, Claudio and Mudry, Christopher},
  title   = {Fractional quantum {Hall} states at zero magnetic field},
  journal = {Phys. Rev. Lett.},
  volume  = {106},
  pages   = {236804},
  year    = {2011},
  doi     = {10.1103/PhysRevLett.106.236804},
}

@article{sheng2011fractional,
  author  = {Sheng, D. N. and Gu, Zheng-Cheng and Sun, Kai and Sheng, L.},
  title   = {Fractional quantum {Hall} effect in the absence of {Landau} levels},
  journal = {Nat. Commun.},
  volume  = {2},
  pages   = {389},
  year    = {2011},
  doi     = {10.1038/ncomms1390},
}

@article{regnault2011fractional,
  author  = {Regnault, Nicolas and Bernevig, B. Andrei},
  title   = {Fractional {Chern} insulator},
  journal = {Phys. Rev. X},
  volume  = {1},
  pages   = {021014},
  year    = {2011},
  doi     = {10.1103/PhysRevX.1.021014},
}

@article{Xie2021MATBGFCI,
  author  = {Xie, Yonglong and Pierce, Andrew T. and Park, Jeong Min and others},
  title   = {Fractional {Chern} insulators in magic-angle twisted bilayer graphene},
  journal = {Nature},
  volume  = {600},
  number  = {7891},
  pages   = {439--443},
  year    = {2021},
  doi     = {10.1038/s41586-021-04002-3},
  url     = {https://doi.org/10.1038/s41586-021-04002-3},
}

@article{Lu2024Pentalayer,
  author  = {Lu, Zhengguang and Han, Tonghang and Yao, Yuxuan and others},
  title   = {Fractional quantum anomalous {Hall} effect in multilayer graphene},
  journal = {Nature},
  volume  = {626},
  number  = {8000},
  pages   = {759--764},
  year    = {2024},
  doi     = {10.1038/s41586-023-07010-7},
  url     = {https://doi.org/10.1038/s41586-023-07010-7},
}

@article{Redekop2024MoTe2,
  author  = {Redekop, Emiel and Xu, Fan and Sun, Zheng and others},
  title   = {Direct magnetic imaging of fractional {Chern} insulators in twisted {MoTe2}},
  journal = {Nature},
  volume  = {635},
  number  = {8039},  pages   = {584--589},
  year    = {2024},
  doi     = {10.1038/s41586-024-08153-x},
  url     = {https://doi.org/10.1038/s41586-024-08153-x},
}

@article{Park2023FQAH,
  author  = {Park, Heonjoon and Cai, Jiaqi and Anderson, Eric and others},
  title   = {Observation of fractionally quantized anomalous {Hall} effect},
  journal = {Nature},
  volume  = {622},
  number  = {7981},
  pages   = {74--79},
  year    = {2023},
  doi     = {10.1038/s41586-023-06536-0},
  url     = {https://doi.org/10.1038/s41586-023-06536-0},
}

@article{Parameswaran2013,
  author  = {Parameswaran, S. A. and Roy, R. and Sondhi, S. L.},
  title   = {Fractional quantum {Hall} physics in topological flat bands},
  journal = {Comptes Rendus Physique},
  volume  = {14},
  number  = {9-10},
  pages   = {816--839},
  year    = {2013},
  doi     = {10.1016/j.crhy.2013.04.002},
}

@article{BergholtzLiu2013,
  author  = {Bergholtz, Emil J. and Liu, Zhao},
  title   = {Topological flat band models and fractional {Chern} insulators},
  journal = {International Journal of Modern Physics B},
  volume  = {27},
  number  = {24},
  pages   = {1330017},
  year    = {2013},
  doi     = {10.1142/S021797921330017X},
}

@misc{BernevigRegnault2012,
      title={Thin-Torus Limit of Fractional Topological Insulators}, 
      author={B. Andrei Bernevig and N. Regnault},
      year={2012},
      eprint={1204.5682},
      archivePrefix={arXiv},
      primaryClass={cond-mat.str-el},
      url={https://arxiv.org/abs/1204.5682}, 
}

@article{Abouelkomsan2020,
  author  = {Abouelkomsan, Ahmed and Liu, Zhao and Bergholtz, Emil J.},
  title   = {Particle-hole duality, emergent Fermi liquids, and fractional {Chern} insulators in moir{\'e} flatbands},
  journal = {Phys. Rev. Lett.},
  volume  = {124},
  pages   = {106803},
  year    = {2020},
  doi     = {10.1103/PhysRevLett.124.106803},
}

@article{RezayiRead2009,
  author  = {Rezayi, Edward H. and Read, Nicholas},
  title   = {Non-{Abelian} quantized {Hall} states of electrons at filling factors 12/5 and 13/5 in the first excited {Landau} level},
  journal = {Phys. Rev. B},
  volume  = {79},
  pages   = {075306},
  year    = {2009},
  doi     = {10.1103/PhysRevB.79.075306},
  url     = {https://doi.org/10.1103/PhysRevB.79.075306},
}

@article{Mong2017Fibonacci,
  author  = {Mong, Roger S. K. and Zaletel, Michael P. and Pollmann, Frank and Papi\'c, Zlatko},
  title   = {Fibonacci anyons and charge density order in the 12/5 and 13/5 quantum {Hall} plateaus},
  journal = {Phys. Rev. B},
  volume  = {95},
  pages   = {115136},
  year    = {2017},
  doi     = {10.1103/PhysRevB.95.115136},
  url     = {https://doi.org/10.1103/PhysRevB.95.115136},
}

@article{Zhu2015NonAbelian,
  author  = {Zhu, W. and Gong, S. S. and Haldane, F. D. M. and Sheng, D. N.},
  title   = {Fractional quantum {Hall} states at $\nu=13/5$ and 12/5 and their non-{Abelian} nature},
  journal = {Phys. Rev. Lett.},
  volume  = {115},
  pages   = {126805},
  year    = {2015},
  doi     = {10.1103/PhysRevLett.115.126805},
  url     = {https://doi.org/10.1103/PhysRevLett.115.126805},
}

@article{LiuLiuBergholtz2024NonAbelian,
  author  = {Liu, Hui and Liu, Zhao and Bergholtz, Emil J.},
  title   = {Non-{Abelian} fractional {Chern} insulators and competing states in flat moir\'e bands},
  journal = {Phys. Rev. Lett.},
  volume  = {135},
  pages   = {106604},
  year    = {2025},
  doi     = {10.1103/43nq-ntqm},
  url     = {https://doi.org/10.1103/43nq-ntqm},
}

@article{jiang2017visualizing,
  title={Visualizing strain-induced pseudomagnetic fields in graphene through an hBN magnifying glass},
  author={Jiang, Y. and Mao, J. and Duan, J. and Lai, X. and Watanabe, K. and Taniguchi, T. and Andrei, E. Y.},
  journal={Nano Letters},
  volume={17},
  pages={2839},
  year={2017}
}

@article{mao2020evidence,
  title={Evidence of flat bands and correlated states in buckled graphene superlattices},
  author={Mao, J. and Milovanovi{\'c}, S. P. and An{\dj}elkovi{\'c}, M. and Lai, X. and Cao, Y. and Watanabe, K. and Taniguchi, T. and Covaci, L. and Peeters, F. M. and Geim, A. K. and Jiang, Y. and Andrei, E. Y.},
  journal={Nature},
  volume={584},
  pages={215--220}, doi={10.1038/s41586-020-2567-3},
  year={2020}
}

@article{forsythe2018band,
  title={Band structure engineering of 2D materials using patterned dielectric superlattices},
  author={Forsythe, C. and Zhou, X. and Watanabe, K. and Taniguchi, T. and Pasupathy, A. and Moon, P. and Koshino, M. and Kim, P. and Dean, C. R.},
  journal={Nature Nanotechnology},
  volume={13},
  pages={566},
  year={2018}
}

@article{shi2019gate,
  title={Gate-tunable flat bands in van der Waals patterned dielectric superlattices},
  author={Shi, L.-k. and Ma, J. and Song, J. C.},
  journal={2D Materials},
  volume={7},
  pages={015028},
  year={2019}
}

@article{PhysRevLett.51.605,
  title = {Fractional Quantization of the Hall Effect: A Hierarchy of Incompressible Quantum Fluid States},
  author = {Haldane, F. D. M.},
  journal = {Phys. Rev. Lett.},
  volume = {51},
  issue = {7},
  pages = {605--608},
  numpages = {0},
  year = {1983},
  month = {8},
  publisher = {American Physical Society},
  doi = {10.1103/PhysRevLett.51.605},
  url = {https://link.aps.org/doi/10.1103/PhysRevLett.51.605}
}

@article{Bernevig_2008,
   title={Model Fractional Quantum Hall States and Jack Polynomials},
   volume={100},
   ISSN={1079-7114},
   url={http://dx.doi.org/10.1103/PhysRevLett.100.246802},
   DOI={10.1103/physrevlett.100.246802}, pages={246802},
   number={24},
   journal={Physical Review Letters},
   publisher={American Physical Society (APS)},
   author={Bernevig, B. Andrei and Haldane, F. D. M.},
   year={2008},
   month={6} }

@article{Ardonne_2008,
   title={Degeneracy of non-Abelian quantum Hall states on the torus: domain walls and conformal field theory},
   volume={2008},
   ISSN={1742-5468},
   url={http://dx.doi.org/10.1088/1742-5468/2008/04/P04016},
   DOI={10.1088/1742-5468/2008/04/p04016},
   number={04},
   journal={Journal of Statistical Mechanics: Theory and Experiment},
   publisher={IOP Publishing},
   author={Ardonne, Eddy and Bergholtz, Emil J and Kailasvuori, Janik and Wikberg, Emma},
   year={2008},
   month=Apr, pages={P04016} }

@article{Jackson_2015,
   title={Geometric stability of topological lattice phases},
   volume={6},
   ISSN={2041-1723},
   url={http://dx.doi.org/10.1038/ncomms9629},
   DOI={10.1038/ncomms9629},
   number={1},
   journal={Nature Communications},
   publisher={Springer Science and Business Media LLC},
   author={Jackson, T. S. and Möller, Gunnar and Roy, Rahul},
   year={2015},
   month={11} }

@article{PhysRevB.90.165139,
  title = {Band geometry of fractional topological insulators},
  author = {Roy, Rahul},
  journal = {Phys. Rev. B},
  volume = {90},
  issue = {16},
  pages = {165139},
  numpages = {7},
  year = {2014},
  month = {10},
  publisher = {American Physical Society},
  doi = {10.1103/PhysRevB.90.165139},
  url = {https://link.aps.org/doi/10.1103/PhysRevB.90.165139}
}

@article{mann2024ai,
  title={Guidelines for ethical use and acknowledgement of large language models in academic writing},
  author={Sebastian Porsdam Mann and Anuraag A. Vazirani and Mateo Aboy and Brian D. Earp and Timo Minssen, I. Glenn Cohen and Julian Savulescu },
  journal={Nat. Mach. Intell.},
  volume={6},
  pages={1272--1274},
  year={2024},
  url={https://www.nature.com/articles/s42256-024-00922-7}
}

@article{6rc3-kjhc,
  title = {Non-Abelian Fibonacci quantum Hall states in tetralayer rhombohedral graphene},
  author = {Timmel, Abigail and Wen, Xiao-Gang},
  journal = {Phys. Rev. B},
  volume = {114}, pages = {125112},
  year = {2026},
  month = {Jul},
  publisher = {American Physical Society},
  doi = {10.1103/6rc3-kjhc},
  url = {https://link.aps.org/doi/10.1103/6rc3-kjhc}
}

@misc{wan2026tunablemultibandgeometryfractional,
      title={Tunable Multiband Geometry and Fractional Phases in Higher Vortexable Systems}, 
      author={Xiaohan Wan and Siddhartha Sarkar and Ting Cao and Mark Rudner and Di Xiao and Kai Sun},
      year={2026},
      eprint={2608.09911},
      archivePrefix={arXiv},
      primaryClass={cond-mat.str-el},
      url={https://arxiv.org/abs/2608.09911}, 
}

@misc{regnault2015entanglementspectroscopyapplicationquantum,
      title={Entanglement Spectroscopy and its Application to the Quantum Hall Effects}, 
      author={N. Regnault},
      year={2015},
      eprint={1510.07670},
      archivePrefix={arXiv},
      primaryClass={cond-mat.str-el},
      url={https://arxiv.org/abs/1510.07670}, 
}

@article{Yang_2025,
   title={Fractional Quantum Anomalous Hall Effect in a Singular Flat Band},
   volume={134},
   ISSN={1079-7114},
   url={http://dx.doi.org/10.1103/PhysRevLett.134.196501},
   DOI={10.1103/physrevlett.134.196501}, pages={196501},
   number={19},
   journal={Physical Review Letters},
   publisher={American Physical Society (APS)},
   author={Yang, Wenqi and Zhai, Dawei and Tan, Tixuan and Fan, Feng-Ren and Lin, Zuzhang and Yao, Wang},
   year={2025},
   month=May }

@misc{butler202613fractionalgaplessinteger,
      title={1/3 Fractional and Gapless Integer Quantum Anomalous Hall States in Rhombohedral Graphene}, 
      author={Jackson P. Butler and Tonghang Han and Andrew DiFabbio and Zach Hadjri and Emily Aitken and Kenji Watanabe and Takashi Taniguchi and Long Ju and Raymond C. Ashoori},
      year={2026},
      eprint={2606.06450},
      archivePrefix={arXiv},
      primaryClass={cond-mat.mes-hall},
      url={https://arxiv.org/abs/2606.06450}, 
}

@book{Giuliani_Vignale_2005, place={Cambridge}, title={Quantum Theory of the Electron Liquid}, publisher={Cambridge University Press}, author={Giuliani, Gabriele and Vignale, Giovanni}, year={2005}}

@article{Simon_2015,
   title={Fractional Chern insulators in bands with zero Berry curvature},
   volume={92},
   ISSN={1550-235X},
   url={http://dx.doi.org/10.1103/PhysRevB.92.195104},
   DOI={10.1103/physrevb.92.195104}, pages={195104},
   number={19},
   journal={Physical Review B},
   publisher={American Physical Society (APS)},
   author={Simon, Steven H. and Harper, Fenner and Read, N.},
   year={2015},
   month=Nov }

@article{PhysRevB.23.5632,
  title = {Quantized Hall conductivity in two dimensions},
  author = {Laughlin, R. B.},
  journal = {Phys. Rev. B},
  volume = {23},
  issue = {10},
  pages = {5632(R)--5633(R)},
  numpages = {0},
  year = {1981},
  month = {May},
  publisher = {American Physical Society},
  doi = {10.1103/PhysRevB.23.5632},
  url = {https://link.aps.org/doi/10.1103/PhysRevB.23.5632}
}

@article{PhysRevLett.84.1535,
  title = {Commensurability, Excitation Gap, and Topology in Quantum Many-Particle Systems on a Periodic Lattice},
  author = {Oshikawa, Masaki},
  journal = {Phys. Rev. Lett.},
  volume = {84},
  issue = {7},
  pages = {1535--1538},
  numpages = {0},
  year = {2000},
  month = {Feb},
  publisher = {American Physical Society},
  doi = {10.1103/PhysRevLett.84.1535},
  url = {https://link.aps.org/doi/10.1103/PhysRevLett.84.1535}
}

@article{boronnitride,
	author = {Dean, C. R. and Young, A. F. and Meric, I. and Lee, C. and Wang, L. and Sorgenfrei, S. and Watanabe, K. and Taniguchi, T. and Kim, P. and Shepard, K. L. and Hone, J.},
	date = {2010/10/01},
	doi = {10.1038/nnano.2010.172},
	id = {Dean2010},
	isbn = {1748-3395},
	journal = {Nature Nanotechnology},
	number = {10},
	pages = {722--726},
	title = {Boron nitride substrates for high-quality graphene electronics},
	url = {https://doi.org/10.1038/nnano.2010.172},
	volume = {5},
	year = {2010}}

@article{Xu2025MultipleChern,
  title = {Multiple Chern Bands in Twisted {MoTe$_2$} and Possible Non-Abelian States},
  author = {Xu, Cheng and Mao, Ning and Zeng, Tiansheng and Zhang, Yang},
  journal = {Physical Review Letters},
  volume = {134},
  pages = {066601},
  year = {2025},
  doi = {10.1103/PhysRevLett.134.066601}
}

@article{Ahn2024NonAbelian,
  title = {Non-Abelian Fractional Quantum Anomalous Hall States and First Landau Level Physics of the Second Moir{\'e} Band of Twisted Bilayer {MoTe$_2$}},
  author = {Ahn, Cheong-Eung and Lee, Wonjun and Yananose, Kunihiro and Kim, Youngwook and Cho, Gil Young},
  journal = {Physical Review B},
  volume = {110},
  pages = {L161109},
  year = {2024},
  doi = {10.1103/PhysRevB.110.L161109}
}

@article{Chen2025RobustNonAbelian,
  title = {Robust Non-Abelian Even-Denominator Fractional Chern Insulator in Twisted Bilayer {MoTe$_2$}},
  author = {Chen, Feng and Luo, Wei-Wei and Zhu, Wei and Sheng, D. N.},
  journal = {Nature Communications},
  volume = {16},
  pages = {2115},
  year = {2025},
  doi = {10.1038/s41467-025-57326-3}
}

@article{Reddy2026AntiTopological,
  title = {Anti-Topological Crystal and Non-Abelian Liquid in Twisted Semiconductor Bilayers},
  author = {Reddy, Aidan P. and Sheng, D. N. and Abouelkomsan, Ahmed and Bergholtz, Emil J. and Fu, Liang},
  journal = {Nature Communications},
  volume = {17},
  pages = {3814},
  year = {2026},
  doi = {10.1038/s41467-026-70916-z}
}

@article{Li2026GeneralizedLandau,
  title = {Abelian and Non-Abelian Fractionalized States in Twisted {MoTe$_2$}: A Generalized Landau-Level Theory},
  author = {Li, Bohao and Ouyang, Yunze and Wu, Fengcheng},
  journal = {Physical Review B},
  volume = {113},
  pages = {195129},
  year = {2026},
  doi = {10.1103/dvry-pfnb}
}

@article{PhysRevLett.127.246403,
  title = {Exact Landau Level Description of Geometry and Interaction in a Flatband},
  author = {Wang, Jie and Cano, Jennifer and Millis, Andrew J. and Liu, Zhao and Yang, Bo},
  journal = {Phys. Rev. Lett.},
  volume = {127},
  issue = {24},
  pages = {246403},
  numpages = {6},
  year = {2021},
  month = {Dec},
  publisher = {American Physical Society},
  doi = {10.1103/PhysRevLett.127.246403},
  url = {https://link.aps.org/doi/10.1103/PhysRevLett.127.246403}
}

@article{oriol2026,
  title = {Refining heuristic predictors of fractional {Chern} insulators using machine learning},
  author = {Mayn\'e i Comas, Oriol and Fonseca, Andr\'e Grossi and Vaidya, Sachin and Solja\ifmmode \check{c}\else \v{c}\fi{}i\ifmmode \acute{c}\else \'{c}\fi{}, Marin},
  journal = {Phys. Rev. B},
  volume = {114},
  issue = {12},
  pages = {125106},
  numpages = {10},
  year = {2026},
  month = {Aug},
  publisher = {American Physical Society},
  doi = {10.1103/qyj5-cd7k},
  url = {https://link.aps.org/doi/10.1103/qyj5-cd7k}
}

@article{Bernevig2012Emergent,
  title     = {Emergent many-body translational symmetries of Abelian and non-Abelian fractionally filled topological insulators},
  author    = {Bernevig, B. Andrei and Regnault, Nicolas},
  journal   = {Physical Review B},
  volume    = {85},
  number    = {7},
  pages     = {075128},
  year      = {2012},
  publisher = {American Physical Society},
  doi       = {10.1103/PhysRevB.85.075128}
}

@article{Liu2013NonAbelian,
  title     = {Non-Abelian fractional Chern insulators from long-range interactions},
  author    = {Liu, Zhao and Bergholtz, Emil J. and Kapit, Eliot},
  journal   = {Physical Review B},
  volume    = {88},
  number    = {20},
  pages     = {205101},
  year      = {2013},
  publisher = {American Physical Society},
  doi       = {10.1103/PhysRevB.88.205101}
}

@article{Zhu2014Identifying,
  title     = {Identifying non-Abelian topological order through minimal entangled states},
  author    = {Zhu, Wei and Gong, Shou-Shu and Haldane, F. D. M. and Sheng, D. N.},
  journal   = {Physical Review Letters},
  volume    = {112},
  number    = {9},
  pages     = {096803},
  year      = {2014},
  publisher = {American Physical Society},
  doi       = {10.1103/PhysRevLett.112.096803}
}

@article{Wang2015Fermionic,
  title     = {Fermionic non-Abelian fractional Chern insulators from dipolar interactions},
  author    = {Wang, Dong and Liu, Zhao and Liu, Wu-Ming and Cao, Junpeng and Fan, Heng},
  journal   = {Physical Review B},
  volume    = {91},
  number    = {12},
  pages     = {125138},
  year      = {2015},
  publisher = {American Physical Society},
  doi       = {10.1103/PhysRevB.91.125138}
}

@article{OzawaMera2021,
  title = {Relations between topology and the quantum metric for {Chern} insulators},
  author = {Ozawa, Tomoki and Mera, Bruno},
  journal = {Phys. Rev. B},
  volume = {104},
  issue = {4},
  pages = {045103},
  year = {2021},
  month = {Jul},
  publisher = {American Physical Society},
  doi = {10.1103/PhysRevB.104.045103},
  url = {https://doi.org/10.1103/PhysRevB.104.045103}
}

@article{MeraOzawa2021,
  title = {K{\"a}hler geometry and {Chern} insulators: Relations between topology and the quantum metric},
  author = {Mera, Bruno and Ozawa, Tomoki},
  journal = {Phys. Rev. B},
  volume = {104},
  issue = {4},
  pages = {045104},
  year = {2021},
  month = {Jul},
  publisher = {American Physical Society},
  doi = {10.1103/PhysRevB.104.045104},
  url = {https://doi.org/10.1103/PhysRevB.104.045104}
}

@article{Kuzmenko2009,
  author  = {Kuzmenko, A. B. and Crassee, I. and
             van der Marel, D. and Blake, P. and Novoselov, K. S.},
  title   = {Determination of the gate-tunable band gap and
             tight-binding parameters in bilayer graphene
             using infrared spectroscopy},
  journal = {Physical Review B},
  volume  = {80},
  number  = {16},
  pages   = {165406},
  year    = {2009},
  doi     = {10.1103/PhysRevB.80.165406}
}

@article{Zhang2008,
  author  = {Zhang, L. M. and Li, Z. Q. and Basov, D. N. and
             Fogler, M. M. and Hao, Z. and Martin, M. C.},
  title   = {Determination of the electronic structure of
             bilayer graphene from infrared spectroscopy},
  journal = {Physical Review B},
  volume  = {78},
  number  = {23},
  pages   = {235408},
  year    = {2008},
  doi     = {10.1103/PhysRevB.78.235408}
}

@article{Zhang2009,
  author  = {Zhang, Yuanbo and Tang, Tsung-Ta and Girit, Caglar and
             Hao, Zhao and Martin, Michael C. and Zettl, Alex and
             Crommie, Michael F. and Shen, Y. Ron and Wang, Feng},
  title   = {Direct observation of a widely tunable bandgap
             in bilayer graphene},
  journal = {Nature},
  volume  = {459},
  number  = {7248},
  pages   = {820--823},
  year    = {2009},
  doi     = {10.1038/nature08105}
}

@article{Kim2023,
  author  = {Kim, Soyun and Kim, Dohun and Watanabe, Kenji and
             Taniguchi, Takashi and Smet, Jurgen H. and Kim, Youngwook},
  title   = {Orbitally Controlled Quantum {Hall} States in
             Decoupled Two-Bilayer Graphene Sheets},
  journal = {Advanced Science},
  volume  = {10},
  number  = {23},
  pages   = {2300574},
  year    = {2023},
  doi     = {10.1002/advs.202300574}
}

@article{Zibrov2017,
  author  = {Zibrov, A. A. and Kometter, C. and Zhou, H. and
             Spanton, E. M. and Taniguchi, T. and Watanabe, K. and
             Zaletel, M. P. and Young, A. F.},
  title   = {Tunable interacting composite fermion phases in a
             half-filled bilayer-graphene {Landau} level},
  journal = {Nature},
  volume  = {549},
  number  = {7672},
  pages   = {360--364},
  year    = {2017},
  doi     = {10.1038/nature23893}
}

@misc{fonseca2026chiralsc,
      title={Zoology of chiral superconductors in {Chern} bands}, 
      author={André Grossi Fonseca and Aidan Reddy and Ahmed Abouelkomsan and Liang Fu and Marin Soljačić},
      year={2026},
      eprint={2609.31839},
      archivePrefix={arXiv},
      primaryClass={cond-mat.str-el},
      url={https://arxiv.org/abs/2609.31839}, 
}
